\documentclass{book}
\usepackage{makeidx,epsfig}
\usepackage{amsthm,amsmath,amssymb}
\usepackage{setspace,graphicx}
\usepackage{Generic}
\usepackage[sort,longnamesfirst]{natbib}

\usepackage{color}

\numberwithin{equation}{section}
\theoremstyle{plain}

\begin{document}
\begin{doublespace} 

%


\chapter[Likelihood-free MCMC]{Likelihood-free Markov chain Monte Carlo}


\begin{center}
\begin{large}
{\em Scott A. Sisson and Yanan Fan}
\end{large}
\end{center}



\section{Introduction}

In Bayesian inference, the posterior distribution for parameters $\theta\in\Theta$ is given by 
$\pi(\theta|y)\propto\pi(y|\theta)\pi(\theta)$, where one's prior beliefs about the unknown parameters, as expressed through the prior distribution $\pi(\theta)$,  is updated by the
observed data $y\in\mathcal{Y}$ via the likelihood function $\pi(y|\theta)$. Inference for
the parameters $\theta$ is then based on the posterior distribution.
Except in simple cases, numerical simulation methods, such as Markov chain Monte Carlo (MCMC),  are required to approximate the integrations needed to summarise features of the posterior
distribution. Inevitably, increasing demands on statistical modelling and computation have resulted in the development of progressively more sophisticated algorithms.

Most recently there has been interest in performing Bayesian analyses for models which are sufficiently complex that the likelihood function  $\pi(y|\theta)$ is either analytically unavailable or computationally prohibitive to evaluate. The classes of algorithms and methods developed to perform Bayesian inference in this setting have become known as {\it likelihood-free computation} or {\it approximate Bayesian computation} \citep{tavare+bgd97,beaumont+zb02,marjoram+mpt03,sisson+ft07,ratmann+ahwr09}. This name refers to the circumventing of explicit evaluation of the likelihood by a simulation-based approximation.

Likelihood-free methods are rapidly gaining popularity as a practical approach to fitting models under the Bayesian paradigm that would otherwise have been computationally impractical. To date they have found widespread usage in a diverse range of applications. These include 
wireless communications engineering \citep{nevat+py08}, quantile distributions \citep{drovandi+p09}, HIV contact tracing \citep{blum+t09}, the evolution of drug resistance in tuberculosis \citep{luciani+sjft09}, population genetics \citep{beaumont+zb02}, protein networks \citep{ratmann+jhsrw07,ratmann+ahwr09}, archaeology \citep{wilkinson+t09}; ecology \citep{jabot+c09}, operational risk \citep{peters+s06}, species migration \citep{hamilton+crhbe05}, chain-ladder claims reserving \citep{peters+fs08}, coalescent models \citep{tavare+bgd97}, $\alpha$-stable models \citep{peters+sf09}, models for extremes \citep{bortot+cs07}, susceptible-infected-removed (SIR) models \citep{toni+wsis09}, pathogen transmission \citep{tanaka+fls06} and human evolution \citep{fagundes+rbnsbe07}.

\begin{table}
  \caption{\label{lfmcmc:table:rejection}The likelihood-free rejection sampling algorithm (Tavar\'e et al., 1997). Accepted parameter vectors are drawn approximately from $\pi(\theta|y)$.}
\begin{center}
\begin{tabular}{l}
  \hline\\
{\bf Likelihood-free rejection sampling algorithm}\\
\hline \\
1. Generate $\theta'\sim\pi(\theta)$ from the prior.\\
2. Generate dataset $x$ from the model $\pi(x|\theta')$.\\
3. Accept $\theta'$ if $x\approx y$.\\
\vspace{1mm}\\
\hline
\end{tabular}
\end{center}
\end{table}

The underlying concept of likelihood-free methods may be simply encapsulated as follows (see Table \ref{lfmcmc:table:rejection}): For a candidate parameter vector $\theta'$, a dataset is generated from the model (i.e. the likelihood function) $x\sim\pi(x|\theta')$. 
If the simulated and observed datasets are similar (in some manner), so that $x\approx y$, then $\theta'$ is a good candidate to have generated the observed data from the given model, and so $\theta'$ is retained and forms
as a part of the samples from the posterior distribution $\pi(\theta |y)$. Conversely, if $x$ and $y$ are dissimilar, then $\theta'$ is unlikely to have generated the observed data for this model, and so $\theta'$ is discarded. The parameter vectors accepted under this approach offer support for $y$ under the model, and so may be considered to be drawn approximately from the posterior distribution $\pi(\theta|y)$. 
In this manner, the evaluation of the likelihood $\pi(y | \theta')$, essential to most Bayesian posterior simulation methods,  is replaced by an estimate of the proximity of a simulated dataset $x\sim\pi(x|\theta')$ to the observed dataset $y$. 
While available in various forms, all likelihood-free methods and models apply this basic principle. 

In this article we aim to provide a tutorial-style exposition of likelihood-free modelling and computation using MCMC simulation. In Section \ref{lfmcmc:section:review} we provide an overview of the models underlying likelihood-free inference, and illustrate the conditions under which these models form an acceptable approximation to the true, but intractable posterior $\pi(\theta|y)$. In Section \ref{lfmcmc:section:samplers} we examine how MCMC-based samplers are able to circumvent evaluation of the intractable likelihood function, while still targetting this approximate posterior model. We also discuss different forms of samplers that have been proposed in order to improve algorithm and inferential performance.
Finally, in Section \ref{lfmcmc:section:guide} we present a step-by-step examination of the various practical issues involved in performing an analysis using likelihood-free methods, 
before concluding with a discussion.

Throughout we assume a basic familiarity with Bayesian inference and the Metropolis-Hastings algorithm. For this relevant background information, the reader is referred to the many useful articles in this volume.

\section{Review of likelihood-free theory and methods}
\label{lfmcmc:section:review}

In this Section we discuss the modelling principles underlying likelihood-free computation.

\subsection{Likelihood-free basics}

A common procedure to improve sampler efficiency in challenging settings is to embed the target posterior within an augmented model. In this setting, auxiliary parameters are introduced into the model whose sole purpose is to facilitate computations (see for example simulated tempering or annealing methods  \citep{geyer+t95,neal03}). Likelihood-free inference adopts a similar approach by augmenting the target posterior from $\pi(\theta|y)\propto\pi(y|\theta)\pi(\theta)$ to
\begin{equation}\label{eqn:LFpost}
	\pi_{LF}(\theta,x|y)\propto\pi(y|x,\theta)\pi(x|\theta)\pi(\theta)
\end{equation}
where the auxiliary parameter $x$ is a (simulated) dataset from $\pi(x|\theta)$ (see Table \ref{lfmcmc:table:rejection}), on the same space as $y\in{\mathcal Y}$ \citep{reeves+p05,wilkinson08}. 
As discussed in more detail below (Section \ref{lfmcmc:sec:nature}), the function $\pi(y|x,\theta)$ is chosen to weight the posterior $\pi(\theta|x)$ with high values in regions where $x$ and $y$ are similar.
The function $\pi(y|x,\theta)$ is assumed to be constant with respect to $\theta$ at the point $x=y$, so that $\pi(y|y,\theta)=c$, for some constant $c>0$, with the result that the target posterior is recovered exactly at $x=y$. That is, $\pi_{LF}(\theta,y|y) \propto \pi(y | \theta)\pi(\theta) $.

Ultimately interest is typically in the marginal posterior
\begin{equation}
\label{lfmcmc:eqn:marginalpost}
	\pi_{LF}(\theta|y)
	\propto
	\pi(\theta)\int_{\mathcal Y}\pi(y|x,\theta)\pi(x|\theta)dx,
\end{equation}
integrating out the auxiliary dataset $x$. The distribution $\pi_{LF}(\theta|y)$ then acts as an approximation to $\pi(\theta|y)$. In practice this integration is performed numerically by simply discarding the realisations of the auxiliary datasets from the output of any sampler targetting the joint posterior $\pi_{LF}(\theta,x|y)$. Other samplers can target $\pi_{LF}(\theta|y)$ directly -- see Section \ref{lfmcmc:sec:marginal}.

\subsection{The nature of the posterior approximation}
\label{lfmcmc:sec:nature}

The likelihood-free posterior distribution $\pi_{LF}(\theta|y)$ will only recover the target posterior $\pi(\theta|y)$ exactly when the function $\pi(y|x,\theta)$ is precisely a point mass at $y=x$ and zero elsewhere \citep{reeves+p05}. In this case
\[
	\pi_{LF}(\theta|y)
	\propto
	\pi(\theta)\int_{\mathcal Y}\pi(y|x,\theta)\pi(x|\theta)dx
	=
	\pi(y|\theta)\pi(\theta).
\]
However, as observed from Table \ref{lfmcmc:table:rejection}, this choice for $\pi(y|x,\theta)$ will result in a rejection sampler with an  acceptance probability  of zero
unless the proposed auxiliary dataset exactly equals the observed data $x=y$. This event will occur with probability zero for all but the simplest applications (involving very low dimensional discrete data). In a similar manner, MCMC-based likelihood-free samplers (Section \ref{lfmcmc:section:samplers}) will also suffer acceptance rates of zero.

In practice, two concessions are made on the form of $\pi(y|x,\theta)$, and each of these can induce some form of approximation into $\pi_{LF}(\theta|y)$ \citep{marjoram+mpt03}.
The first allows the function to be a standard smoothing kernel density, $K$, centered at the point $x=y$ and with scale determined by a parameter vector $\epsilon$, usually taken as a scalar. In this manner
\[
	\pi_\epsilon(y|x,\theta) = \frac{1}{\epsilon}K\left(\frac{|x-y|}{\epsilon}\right)
\]
weights the intractable likelihood with high values in regions $x\approx y$ where the auxiliary and observed datasets are similar, and with low values in regions where they are not similar \citep{beaumont+zb02,blum09,peters+fs08}. The interpretation of likelihood-free models in the non-parametric framework is of current research interest \citep{blum09}.

The second concession on the form of $\pi_\epsilon(y|x,\theta)$ permits the comparison of the datasets, $x$ and $y$, to occur through a low-dimensional vector of summary statistics $T(\cdot)$, where $\dim(T(\cdot))\geq\dim(\theta)$. Accordingly, given the improbability of generating an auxiliary dataset such that $x\approx y$, the function 
\begin{equation}
\label{lfmcmc:eqn:kernel}
	\pi_\epsilon(y|x,\theta) = \frac{1}{\epsilon}K\left(\frac{|T(x)-T(y)|}{\epsilon}\right)
\end{equation}
will provide regions of high value when $T(x)\approx T(y)$ and low values otherwise. If the vector of summary statistics is also sufficient for the parameters $\theta$, then comparing the summary statistics of two datasets will be equivalent to comparing the datasets themselves. Hence there will be no loss of information in model fitting, and accordingly no further approximation will be introduced into $\pi_{LF}(\theta|y)$. However, the event $T(x)\approx T(y)$ will be substantially more likely than $x\approx y$, and so likelihood-free samplers based on summary statistics $T(\cdot)$ will in general be considerably more efficient in terms of acceptance rates than those based on full datasets \citep{tavare+bgd97,pritchard+spf99}. As noted by \citet{mckinley+cd09}, the procedure of model fitting via summary statistics $T(\cdot)$ permits the application of likelihood-free inference in situations where the observed data $y$ are incomplete. 

Note that under the form (\ref{lfmcmc:eqn:kernel}), $\lim_{\epsilon\rightarrow 0}\pi_\epsilon(y|x,\theta)$ is a point mass on $T(x)=T(y)$. Hence, if $T(\cdot)$ are also sufficient statistics for $\theta$, then $\lim_{\epsilon\rightarrow 0}\pi_{LF}(\theta|y)=\pi(\theta|y)$ exactly recovers the intractable posterior \citep{reeves+p05}.
Otherwise, if $\epsilon>0$ or if $T(\cdot)$ are not sufficient statistics, then the likelihood-free approximation to $\pi(\theta|y)$ is given by $\pi_{LF}(\theta|y)$ in 
(\ref{lfmcmc:eqn:marginalpost}).

A frequently utilised weighting function $\pi_\epsilon(y|x,\theta)$ is the uniform kernel density \citep{marjoram+mpt03,tavare+bgd97}, whereby $T(y)$ is uniformly distributed on the sphere centered at $T(x)$ with radius $\epsilon$. This is commonly written as
\begin{equation}
\label{lfmcmc:eqn:uniform}
	\pi_\epsilon(y|x,\theta)\propto\left\{
	\begin{array}{ll}1&\mbox{if }\rho(T(x),T(y))\leq\epsilon\\0&\mbox{otherwise}\end{array}
	\right.
\end{equation}
where $\rho$ denotes a distance measure (e.g. Euclidean) between $T(x)$ and $T(y)$. In the form of (\ref{lfmcmc:eqn:kernel}) this is expressed as $\pi_\epsilon(y|x,\theta)=\epsilon^{-1}K_u(\rho(T(x),T(y))/\epsilon)$, where $K_u$ is the uniform kernel density.
Alternative kernel densities that have been implemented include the Epanechnikov kernel \citep{beaumont+zb02}, a non-parametric density estimate \citep{ratmann+ahwr09} (see Section \ref{lfmcmc:sec:epsaug}),
and
the Gaussian kernel density \citep{peters+fs08}, whereby $\pi_\epsilon(y|x,\theta)$ is centered at $T(x)$ and scaled by $\epsilon$, so that
$T(y)\sim N(T(x),\Sigma\epsilon^2)$ for some covariance matrix $\Sigma$.

\subsection{A simple example}
\label{lfmcmc:sec:example}

As an illustration, we examine the deviation of the likelihood-free approximation from the target posterior in a simple example. Consider the case where $\pi(\theta|y)$ is the univariate $N(0,1)$ density. To realise this posterior in the likelihood-free setting, we specify the likelihood as $x\sim N(\theta,1)$, define $T(x)=x$ as a sufficient statistic for $\theta$ (the sample mean) and set the observed data $y=0$. With the prior $\pi(\theta)\propto 1$ for convenience, if the weighting function $\pi_\epsilon(y|x,\theta)$ is given by (\ref{lfmcmc:eqn:uniform}), with $\rho(T(x),T(y))=|x-y|$, or if $\pi_\epsilon(y|x,\theta)$ is a Gaussian density with $y\sim N(x,\epsilon^2/3)$ then respectively
\[
	\pi_{LF}(\theta|y)\propto\frac{\Phi(\epsilon-\theta)-\Phi(-\epsilon-\theta)}{2\epsilon}
	\qquad
	\mbox{and}
	\qquad
	\pi_{LF}(\theta|y)=N(0,1+\epsilon^2/3),
\]
where $\Phi(\cdot)$ denotes the standard Gaussian cumulative distribution function. The factor of 3 in the Gaussian kernel density ensures that both uniform and Gaussian kernels have the same standard deviation. In both cases $\pi_{LF}(\theta|y)\rightarrow N(0,1)$ as $\epsilon\rightarrow 0$.

\begin{figure}
\begin{center}
\rotatebox{0}{\includegraphics[width=16cm]{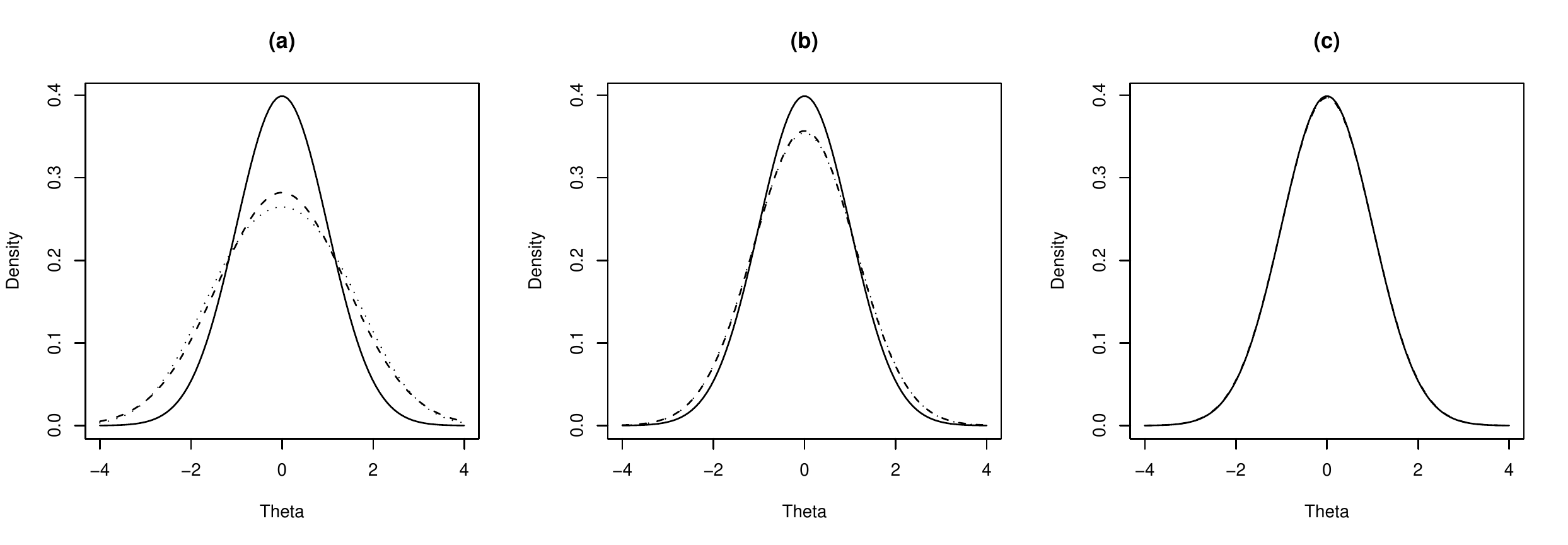}} 
 \caption{\label{lfmcmc:fig:toy}Comparison of likelihood-free approximations to the $N(0,1)$ target posterior (solid line). Likelihood-free posteriors are constructed using uniform (dotted line) and Gaussian (dashed line) kernel weighting densities $\pi_\epsilon(y|x,\theta)$. Panels (a)--(c) correspond to $\epsilon$ values of  $\sqrt{3}$, $\sqrt{3}/2$ and $\sqrt{3}/10$ 
 respectively.}
\end{center}
\end{figure}

The two likelihood-free approximations are illustrated in Figure \ref{lfmcmc:fig:toy} which compares the target $\pi(\theta|y)$ to both forms of $\pi_{LF}(\theta|y)$ for different values of $\epsilon$. Clearly, as $\epsilon$ gets smaller then $\pi_{LF}(\theta|y)\approx\pi(\theta|y)$ becomes a better approximation. Conversely, as $\epsilon$ increases, then so does the posterior variance in the likelihood-free approximation. There is only a small difference between using uniform and Gaussian weighting functions in this case.

Suppose now that an alternative vector of summary statistics $\tilde{T}(\cdot)$ also permits unbiased estimates of $\theta$, but is less efficient than $T(\cdot)$, 
with a relative efficiency of $e\leq1$. As noted by A. N.  Pettitt (personal communication), for the above example with the Gaussian kernel density for $\pi_\epsilon(y|x,\theta)$, the likelihood-free approximation using $\tilde{T}(\cdot)$ becomes $\pi_{LF}(\theta|y)=N(0,1/e+\epsilon^2/3)$. 
The $1/e$ term can easily be greater than the $\epsilon^2/3$ term, especially as practical interest is in small $\epsilon$.
This example illustrates that inefficient statistics can often determine the quality of the posterior approximation, and that this approximation can remain poor even for $\epsilon= 0$.

Accordingly, it is common in practice to aim to reduce $\epsilon$ as low as is computationally feasible.
However, in certain circumstances, it is not clear that doing so 
will result in a better approximation to $\pi(\theta|y)$ than for a larger $\epsilon$. This point is illustrated in Section \ref{lfmcmc:section:ss}.

\section{Likelihood-free MCMC samplers}
\label{lfmcmc:section:samplers}

A Metropolis-Hastings sampler may be constructed to target the augmented likelihood-free posterior  $\pi_{LF}(\theta,x|y)$ (given by \ref{eqn:LFpost}) 
without directly evaluating the intractable likelihood \citep{marjoram+mpt03}. Consider a proposal distribution for this sampler with the factorisation
\[
	q[(\theta,x),(\theta',x')] =q(\theta,\theta') \pi(x'|\theta').
\]
That is, when at a current algorithm state $(\theta,x)$, a new parameter vector $\theta'$ is drawn from a proposal distribution $q(\theta,\theta')$, 
and conditionally on $\theta'$ a proposed dataset $x'$ is generated from the model $x'\sim\pi(x|\theta')$. 
Following standard arguments, to achieve a Markov chain with stationary distribution $\pi_{LF}(\theta,x|y)$, we enforce
the detailed-balance (time-reversibility) condition 
\begin{equation}\label{eqn:revcond}
\pi_{LF}(\theta, x|y)P[(\theta, x),(\theta', x')]  = \pi_{LF}(\theta',x'|y)P[(\theta', x'),(\theta, x)] 
\end{equation}
where
the Metropolis-Hastings transition probability is given by
\[
	P[(\theta, x),(\theta', x')]=q[(\theta, x),(\theta', x')]\alpha[(\theta, x),(\theta', x')].
\]
The probability of accepting a move from $(\theta,x)$ to $(\theta',x')$  within the Metropolis-Hastings framework is then given by $\min\{1,\alpha[(\theta,x),(\theta',x')]\}$, where
\begin{eqnarray}
\label{lfmcmc:eqn:acc}
	\alpha[(\theta,x),(\theta',x')]
	&=&
	\frac{\pi_{LF}(\theta',x'|y)q[(\theta',x'),(\theta,x)]}{\pi_{LF}(\theta,x|y)q[(\theta,x),(\theta',x')]}\nonumber\\
	&=&
	\frac{\pi_\epsilon(y|x',\theta')\pi(x'|\theta')\pi(\theta')}{\pi_\epsilon(y|x,\theta)\pi(x|\theta)\pi(\theta)}\frac{q(\theta',\theta) \pi(x|\theta)}{q(\theta,\theta') \pi(x'|\theta')}\\
	&=&
	\frac{\pi_\epsilon(y|x',\theta')\pi(\theta')q(\theta',\theta)}{\pi_\epsilon(y|x,\theta)\pi(\theta)q(\theta,\theta')}.\nonumber
\end{eqnarray}
Note that the intractable likelihoods 
do not need to be evaluated in the acceptance probability calculation (\ref{lfmcmc:eqn:acc}), leaving a computationally tractable expression which can now be evaluated. 
Without loss of generality we may assume that $\min\{1,\alpha[(\theta',x'),(\theta,x)]\}=1$, and hence the
detailed-balance condition (\ref{eqn:revcond}), is satisfied since
\begin{eqnarray*}
 \pi_{LF}(\theta, x|y)P[(\theta, x),(\theta', x')]
& =& \pi_{LF}(\theta, x|y)q[(\theta, x),(\theta', x')]\alpha[(\theta, x),(\theta', x')]\\
&=& \frac{\pi_{LF}(\theta, x|y)q(\theta,\theta')\pi(x'|\theta')\pi_\epsilon(y|x',\theta')\pi(\theta')q(\theta',\theta)}{\pi_\epsilon(y|x,\theta)\pi(\theta)q(\theta,\theta')}\\
&=&\frac{\pi_\epsilon(y|x,\theta)\pi(x|\theta)\pi(\theta)q(\theta,\theta')\pi(x'|\theta')\pi_\epsilon(y|x',\theta')\pi(\theta')q(\theta',\theta)}{\pi_\epsilon(y|x,\theta)\pi(\theta)q(\theta,\theta')}\\
&=&\pi_\epsilon(y|x',\theta')\pi(x'|\theta')\pi(\theta')q(\theta',\theta)\pi(x|\theta)\\
&=&\pi_{LF}(\theta',x'|y)P[(\theta', x'),(\theta, x)].
\end{eqnarray*}

\begin{table}
  \caption{\label{lfmcmc:table:algorithm}The likelihood-free MCMC algorithm, generalised from \cite{marjoram+mpt03}.}
\begin{center}
\begin{tabular}{ll}
  \hline\\
  \multicolumn{2}{l}{\bf LF-MCMC Algorithm}\\ 
\hline\\
1. & Initialise $(\theta_0,x_0)$ and $\epsilon$. Set $t=0$.\\
\vspace{0.5mm}\\
\multicolumn{2}{l}{At step $t$:}\\
2. & Generate $\theta'\sim q(\theta_t,\theta)$ from a proposal distribution.\\
3. & Generate $x'\sim\pi(x|\theta')$ from the model given $\theta'$.\\
4. & With probability $\min\{1, \frac{\pi_\epsilon(y|x',\theta')\pi(\theta')q(\theta',\theta_t)}{\pi_\epsilon(y|x_t,\theta_t)\pi(\theta_t)q(\theta_t,\theta')}\}$\ set $(\theta_{t+1},x_{t+1})=(\theta',x')$\\
& otherwise set $(\theta_{t+1},x_{t+1})=(\theta_t,x_t)$.\\
5. & Increment $t=t+1$ and go to 2.\\
\vspace{1mm}\\
\hline\\
\end{tabular}
\end{center}
\end{table}

The MCMC algorithm targetting $\pi_{LF}(\theta,x|y)$, adapted from \cite{marjoram+mpt03}, is listed in Table \ref{lfmcmc:table:algorithm}. 
The sampler generates the Markov chain sequence $(\theta_t,x_t)$ for $t\geq0$, although in practice, it is only necessary to store the vectors of summary statistics $T(x_t)$ and $T(x')$ at any stage in the algorithm. This is particularly useful when the auxiliary datasets $x_t$ are large and complex.

An interesting feature of this sampler is that its acceptance rate is directly related to the value of the true likelihood function $\pi(y|\theta')$ at the proposed vector $\theta'$ \citep{sisson+ft07}. This is most obviously seen when using the uniform kernel weighting function (\ref{lfmcmc:eqn:uniform}), as proposed moves to $(\theta',x')$ can only be accepted if $\rho(T(x'),T(y))\leq\epsilon$, and this occurs with a probability in proportion to the likelihood. For low $\epsilon$ values this can result in very low acceptance rates, particularly in the tails of the distribution, thereby affecting chain mixing in regions of low posterior density. See Section \ref{lfmcmc:sec:mixing} for an illustration. However the LF-MCMC algorithm offers improved acceptance rates over rejection sampling-based likelihood-free algorithms \citep{marjoram+mpt03}.

We now examine a number of variations on the basic LF-MCMC algorithm which have been proposed either to improve sampler performance, or to examine model goodness-of-fit.

\subsection{Marginal space samplers}
\label{lfmcmc:sec:marginal}

Given the definition of $\pi_{LF}(\theta|y)$ in (\ref{lfmcmc:eqn:marginalpost}), an unbiased pointwise estimate of the marginal posterior distribution is available through Monte Carlo integration as
\begin{equation}
\label{lfmcmc:eqn:MC}
	\pi_{LF}(\theta|y) \approx \frac{\pi(\theta)}{S}\sum_{s=1}^S\pi_\epsilon(y|x^s,\theta)
\end{equation}
where $x^1,\ldots,x^S$ are independent draws from the model $\pi(x|\theta)$ \citep{marjoram+mpt03, peters+fs08,reeves+p05,ratmann+ahwr09,toni+wsis09,sisson+ft07,wegmann+le09}. This then permits an MCMC sampler to be constructed directly targetting the likelihood-free marginal posterior $\pi_{LF}(\theta|y)$. In this setting, the probability of accepting a proposed move from $\theta$ to $\theta'\sim q(\theta,\theta')$ is given by $\min\{1, \alpha(\theta,\theta')\}$ where
\begin{equation}
\label{lfmcmc:eqn:montecarlo}
	\alpha(\theta,\theta')
	=
	\frac{\pi_{LF}(\theta'|y)q(\theta',\theta)}{\pi_{LF}(\theta|y)q(\theta,\theta')}
	\approx
	\frac{\frac{1}{S}\sum_s\pi_\epsilon(y|{x'}^s,\theta')\pi(\theta')q(\theta',\theta)}{\frac{1}{S}\sum_s\pi_\epsilon(y|x^s,\theta)\pi(\theta)q(\theta,\theta')}
\end{equation}
where ${x'}^1,\ldots,{x'}^S\sim\pi(x|\theta')$. As the Monte Carlo approximation (\ref{lfmcmc:eqn:MC}) becomes more accurate as $S$ increases, the performance and acceptance rate of the marginal likelihood-free sampler will gradually approach that of the equivalent standard MCMC sampler.

However, the above ratio of two unbiased likelihood estimates is only unbiased as $S\rightarrow\infty$. Hence, the above sampler will only approximately target $\pi_{LF}(\theta|y)$ for large $S$, which makes it highly inefficient. However, note that estimating $\alpha(\theta,\theta')$ with $S=1$ exactly recovers (\ref{lfmcmc:eqn:acc}), the acceptance probability of the MCMC algorithm targetting $\pi_{LF}(\theta,x|y)$. That is, the marginal space likelihood-free sampler 
with $S=1$
is precisely the likelihood-free MCMC sampler in Table \ref{lfmcmc:table:algorithm}. As the sampler targetting $\pi_{LF}(\theta,x|y)$ also provides unbiased estimates of the marginal $\pi_{LF}(\theta|y)$, it follows that
the likelihood-free sampler targetting $\pi_{LF}(\theta|y)$ directly is also unbiased in practice  \citep{sisson+pfb08}.
A similar argument for $S>1$ can also be made, as outlined below.

An alternative augmented likelihood-free posterior distribution is given by
\begin{eqnarray*}
	\pi_{LF}(\theta,x_{1:S}|y)& \propto & \pi_\epsilon(y|x_{1:S},\theta)\pi(x_{1:S}|\theta)\pi(\theta)\\
	&  := & 
	\left[\frac{1}{S}\sum_{s=1}^S\pi_\epsilon(y|x^s,\theta)\right]
	\left[\prod_{s=1}^S\pi(x^s|\theta)]\right]\pi(\theta),
\end{eqnarray*}
where  $x_{1:S}=(x^1,\ldots,x^S)$ represents $s=1,\ldots,S$ replicate auxiliary datasets $x^s\sim\pi(x|\theta)$.
This posterior, generalised from \citet{delmoral+dj08}, is based on the more general expected auxiliary variable approach of  \citet{andrieu+bdr08}, where the summation form of $\pi_\epsilon(y|x_{1:S},\theta)$ describes this expectation. 
The resulting marginal posterior $\pi_{LF}^S(\theta|y)=\int_{{\mathcal Y}^S}\pi_{LF}(\theta,x_{1:S},\theta|y)dx_{1:S}$ is the same for all $S$, namely $\pi_{LF}^S(\theta|y)=\pi_{LF}(\theta|y)$.

The motivation for this form of posterior is that that a sampler targetting $\pi_{LF}(\theta,x_{1:S}|y)$, for $S>1$, will possess improved sampler performance  compared to an equivalent sampler targetting $\pi_{LF}(\theta,x|y)$, through a reduction in the variability of the Metropolis-Hastings acceptance probability.
With the natural
choice of proposal density given by
\[
	q[(\theta,x_{1:S}),(\theta',x'_{1:S})]=q(\theta,\theta')\prod_{s=1}^S\pi(x'^s|\theta'),
\]
where $x'_{1:S}=(x'^1,\ldots,x'^S)$, the acceptance probability of a Metropolis-Hastings algorithm targetting $\pi_{LF}(\theta,x_{1:S}|y)$ reduces to
\begin{equation}
\label{lfmcmc:eqn:exact}
	\alpha[(\theta,x_{1:S}),(\theta',x'_{1:S})]
	=
	\frac{\frac{1}{S}\sum_s\pi_\epsilon(y|{x'}^s,\theta')\pi(\theta')q(\theta',\theta)}{\frac{1}{S}\sum_s\pi_\epsilon(y|x^s,\theta),\pi(\theta)q(\theta,\theta')}.
\end{equation}
This is the same acceptance probability
(\ref{lfmcmc:eqn:montecarlo})
as a marginal likelihood-free sampler targetting $\pi_{LF}(\theta|y)$ directly, using $S$ Monte Carlo draws to estimate $\pi_{LF}(\theta|y)$ pointwise, via (\ref{lfmcmc:eqn:MC}).
Hence, both marginal and augmented likelihood-free samplers possess identical mixing and efficiency properties. The difference between the two is that the marginal sampler acceptance probability (\ref{lfmcmc:eqn:montecarlo}) is approximate for finite $S$, whereas the augmented sampler acceptance probability (\ref{lfmcmc:eqn:exact}) is exact. However,  clearly the marginal likelihood-free sampler is, in practice, unbiased for all $S\geq 1$.
See \cite{sisson+pfb08} a for more detailed analysis.

\subsection{Error-distribution augmented samplers}
\label{lfmcmc:sec:epsaug}

In all likelihood-free MCMC algorithms, low values of $\epsilon$ result in slowly mixing chains through low acceptance rates. However, it also provides a potentially more accurate posterior approximation $\pi_{LF}(\theta|y)\approx\pi(\theta|y)$. Conversely, MCMC samplers with larger $\epsilon$ values may possess improved chain mixing and efficiency, although at the expense of a poorer posterior approximation (e.g. Figure \ref{lfmcmc:fig:toy}).
Motivated by a desire for improved sampler efficiency while realising low $\epsilon$ values, \citet{bortot+cs07} proposed augmenting the likelihood-free posterior approximation to include $\epsilon$, so that
\[
	\pi_{LF}(\theta, x, \epsilon|y)\propto\pi_\epsilon(y|x,\theta)\pi(x|\theta)\pi(\theta)\pi(\epsilon).
\]
Accordingly, $\epsilon$ is treated as a tempering parameter in the manner of simulated tempering \citep{geyer+t95}, with larger and smaller values respectively corresponding to ``hot'' and ``cold'' tempered posterior distributions. The density $\pi(\epsilon)$ is a pseudo-prior, which serves only to influence the mixing of the sampler through the tempered distributions. 
\citet{bortot+cs07} suggested using a distribution which favours small $\epsilon$ values for accuracy, while permitting large values to improve chain acceptance rates.
The approximation to the true posterior $\pi(\theta|y)$ is then given by
\[
	\pi_{LF}^{\mathcal E}(\theta|y)=\int_{\mathcal E}\int_{\mathcal Y}\pi_{LF}(\theta,x,\epsilon|y)dxd\epsilon
\]
where $\epsilon\in{\mathcal E}\subseteq{\mathbb R}^+$. Sampler performance aside, this approach permits an {\it a posteriori} evaluation of an appropriate value $\epsilon=\epsilon^*$ such that $\pi_{LF}^{\mathcal E}(\theta|y)$ with ${\mathcal E}=[0,\epsilon^*]$ provides an acceptable approximation to $\pi(\theta|y)$.

An alternative error-distribution augmented model was proposed by \cite{ratmann+ahwr09} with the aim of diagnosing model mis-specification for the observed data $y$. For the vector of summary statistics $T(x)=(T_1(x),\ldots,T_R(x))$, the discrepancy between the model $\pi(x|\theta)$ and the observed data is given by $\tau=(\tau_1,\ldots,\tau_R)$, where $\tau_r=T_r(x)-T_r(y)$, for $r=1,\ldots,R$, is the error under the model in reproducing the $r$-th element of $T(\cdot)$.
The joint distribution of model parameters and model errors is defined as
\begin{eqnarray}
\label{lfmcmc:eqn:ratmann}
	\pi_{LF}(\theta, x_{1:S}, \tau|y)
	& \propto &
	\pi_\epsilon(y|\tau,x_{1:S}, \theta) \pi(x_{1:S}|\theta) \pi(\theta)\pi(\tau)\nonumber\\
	& := &
	\min_r \hat{\xi}_r(\tau_r|y, x_{1:S}, \theta) \pi(x_{1:S}|\theta) \pi(\theta)\pi(\tau),
\end{eqnarray}
where the univariate error distributions
\begin{equation}
\label{lfmcmc:eqn:ratmannK}
	\hat{\xi}_r(\tau_r|y, x_{1:S}, \theta) =\frac{1}{S\epsilon_r}\sum_{s=1}^SK\left(\frac{\tau_r-\left[T_r(x^s) -T_r(y)\right]}{\epsilon_r}\right)
\end{equation}
are constructed from smoothed kernel density estimates of model errors, estimated from $S$ auxiliary datasets $x^1,\ldots,x^S$,  and where $\pi(\tau)=\prod_r\pi(\tau_r)$, the joint prior distribution for the model errors, is centered on zero, reflecting that the model is assumed plausible {\it a priori}.
The terms $\min_r \hat{\xi}_r(\tau_r|y, x, \theta)$ and $\pi(\tau)$ take the place of the  weighting function $\pi_\epsilon(y|\tau,x_{1:S},\theta)$.
The minimum of the univariate densities $\hat{\xi}_r(\tau_r|y, x, \theta)$ is taken over the $R$ model errors to reflect the most conservative estimate of model adequacy, while also reducing the computation on the multivariate $\tau$ to its univariate component margins. The smoothing bandwidths $\epsilon_r$ of each summary statistic $T_r(\cdot)$ are dynamically estimated during sampler implementation as twice the interquartile range of $T_r(x^s) -T_r(y)$, given $x^1,\ldots,x^S$.

Assessment of model adequacy can then be based on $\pi_{LF}(\tau|y)=\int_{\Theta}\int_{{\mathcal Y}^S}\pi_{LF}(\theta,x_{1:S},\tau|y)dx_{1:S}d\theta$, the posterior distribution of the model errors. If the model is adequately specified then $\pi_{LF}(\tau|y)$ should be centered on the zero vector. If this is not the case then the model is mis-specified.
The nature of the departure of $\pi_{LF}(\tau|y)$ from the origin e.g. via one or more summary statistics $T_r(\cdot)$, may indicate the manner in which the model is deficient.
See e.g. \citet{wilkinson08} for further assessment of model errors in likelihood-free models.

\subsection{Potential alternative MCMC samplers}
\label{lfmcmc:sec:alternative}

Given the variety of MCMC techniques available for standard Bayesian inference, there are a number of currently unexplored ways in which these might be adapted to improve the performance of likelihood-free MCMC samplers.

For example, within the class of marginal space samplers (Section \ref{lfmcmc:sec:marginal}), the number of Monte Carlo draws  $S$ determines the quality of the estimate of $\pi_{LF}(\theta|y)$ (c.f. \ref{lfmcmc:eqn:MC}). A standard implementation of the delayed-rejection algorithm \citep{tierney+m99}
would permit rejected proposals based on poor but computationally cheap posterior estimates (i.e. using low-moderate $S$), to generate more accurate but computationally expensive second-stage proposals (using large $S$), thereby adapting the computational overheads of the sampler to the required performance.

Alternatively, coupling two or more Markov chains targetting $\pi_{LF}(\theta,x|y)$, each utilising a different $\epsilon$ value, would achieve improved mixing in the ``cold'' distribution (i.e. the chain with the lowest $\epsilon$) through the switching of states between neighbouring (in an $\epsilon$ sense) chains \citep{pettitt06}. This could be particularly useful in multi-modal posteriors. 
While this flexibility is already available with continuously varying $\epsilon$ in the augmented sampler targetting $\pi_{LF}(\theta,x,\epsilon|y)$ (\citet{bortot+cs07}, Section \ref{lfmcmc:sec:epsaug}), there are benefits to constructing samplers from multiple chain sample-paths.

Finally, likelihood-free MCMC samplers have to date focused on tempering distributions based on varying $\epsilon$. While not possible in all applications, there is clear scope for a class of algorithms based on tempering on the number of observed datapoints from which the summary statistics $T(\cdot)$ are calculated.
Lower numbers of datapoints  will produce greater variability in the summary statistics, in turn generating wider posteriors for the parameters $\theta$, but with lower
computational overheads required to generate the auxiliary data $x$.

\section{A practical guide to likelihood-free MCMC}
\label{lfmcmc:section:guide}

In this Section we examine various practical aspects of likelihood-free computation under a simple worked analysis.
For observed data $y=(y_1,\ldots,y_{20})$ consider two candidate models: $y_i\sim\mbox{Exponential}(\lambda)$ and $y_i\sim\mbox{Gamma}(k,\psi)$, where model equivalence is obtained under $k=1,\psi=1/\lambda$. Suppose that the sample mean and standard deviation of $y$ are available as summary statistics
$T(y)=(\bar{y},s_y)=(4,1)$, and that interest is in fitting each model and in establishing model adequacy. Note that the summary statistics $T(\cdot)$ are sufficient for $\lambda$ but not for $(k,\psi)$, where they form moment-based estimators.
For the following we consider flat priors $\pi(\lambda)\propto 1$, $\pi(k,\psi)\propto 1$ for convenience. The true posterior distribution under the Exponential($\lambda$) model is $\lambda|y\sim\mbox{Gamma}(21,80)$.

\subsection{An exploratory analysis}
\label{lfmcmc:sec:explore}

\begin{figure}
\begin{center}
\rotatebox{0}{\includegraphics[width=16cm]{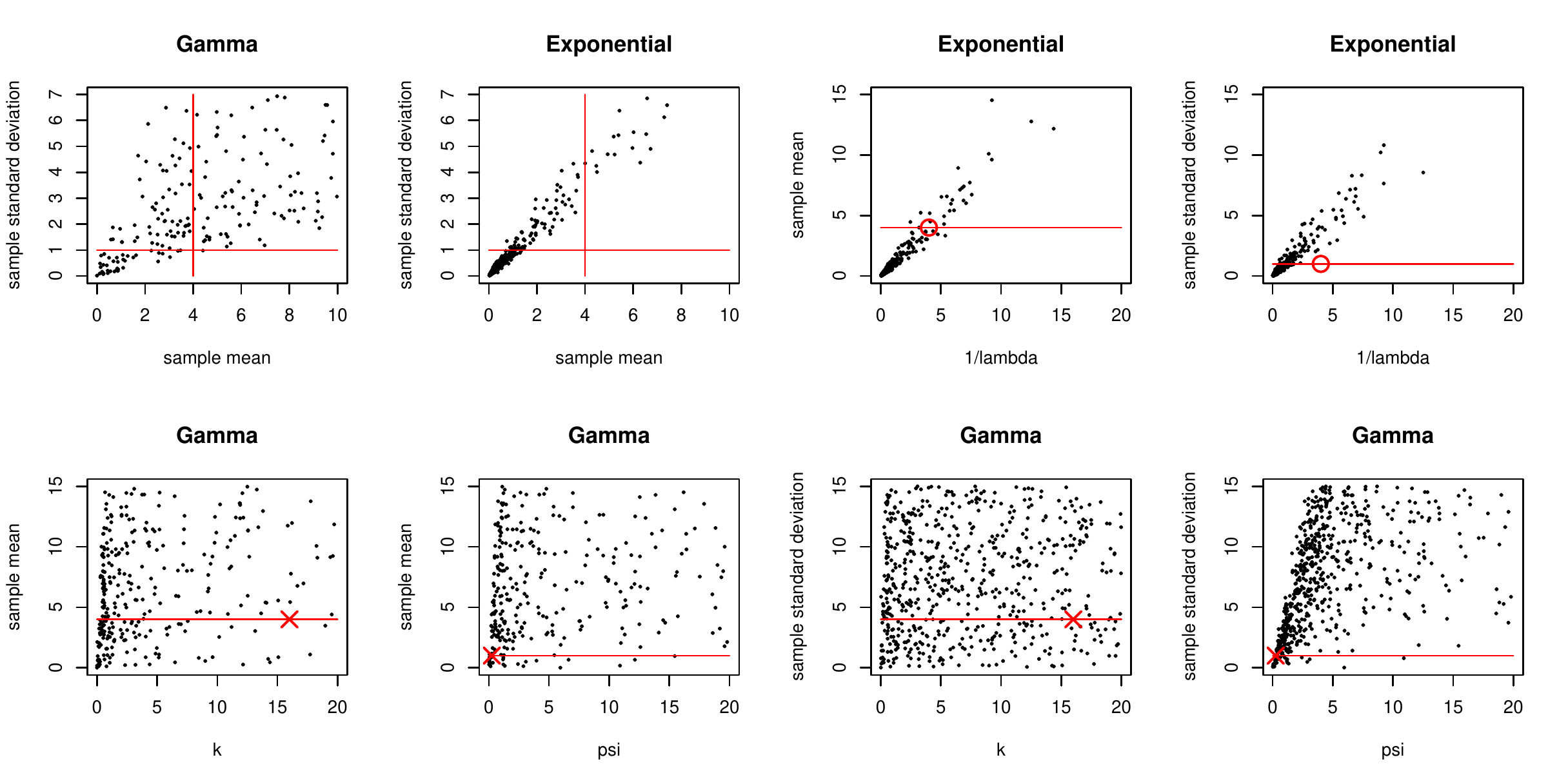}} 
 \caption{\label{lfmcmc:fig:sim:fig1}Scatterplots of summary statistics $T(x)=(\bar{x},s_x)$ and parameter values $\lambda,k,\psi$ under both Exponential$(\lambda)$ and Gamma$(k,\psi)$ models, based on 2000 realisations $\lambda,k,\psi\sim U(0,20)$. Horizontal and vertical lines denote observed summary statistics $T(y)=(4,1)$. Circles denote the MLE of $\hat{\lambda}=1/\bar{y}=1/4$ under the Exponential model. Crosses denote method of moments estimators $\hat{k}=\bar{y}^2/s_y^2=16$ and $\hat{\psi}=s_y^2/\bar{y}=1/4$ under the Gamma model.
}
\end{center}
\end{figure}

An initial exploratory investigation of model adequacy is illustrated in Figure \ref{lfmcmc:fig:sim:fig1}, 
which presents scatterplots of summary statistics versus summary statistics, and summary statistics versus parameter values under each model. Images are based on 2000 parameter realisations $\lambda,k,\psi\sim U(0,20)$ followed by summary statistic generation under each model parameter. Horizontal and vertical lines denote the values of the observed summary statistics $T(y)$.

From the plots of sample means against standard deviations, $T(y)$ is clearly better represented by the Gamma than the Exponential model. The observed summary statistics (i.e. the intersection of horizontal and vertical lines) lie in regions of relatively lower prior predictive density under the Exponential model, compared to the Gamma. That is, {\it a priori}, the statistics $T(y)$ appear more probable under the more complex model.

Consider the plots of $\lambda^{-1}$ versus $T(x)$ under the Exponential model. The observed statistics $T(y)$ individually impose competing requirements on the Exponential parameter. An observed sample mean of $\bar{y}=4$ indicates that $\lambda^{-1}$  is most likely in the approximate range $[3,5]$ (indicated by those $\lambda^{-1}$ values where the horizontal line intersects with the density). However, the sample standard deviation $s_y=1$ independently suggests that  $\lambda^{-1}$ is most likely in the approximate range $[0.5,1.5]$. 
If either $\bar{x}$ or $s_x$ were the only summary statistic, then only one of these ranges are appropriate, and the observed data would be considerably more likely under the Exponential model. 
However, the relative model fits and model adequacies of the Exponential and Gamma 
can only be evaluated by using the same summary statistics on each model. (Otherwise, the model with the smaller number of summary statistics will be considered the most likely model, simply because it is more probable to match fewer statistics.)
As a result, the competing constraints on $\lambda$ through the statistics $\bar{x}$ and $s_y$  are so jointly improbable under the Exponential model that simulated and observed data will rarely coincide, making $T(y)$ very unlikely under this model. This is a strong indicator of model inadequacy.

In contrast, the plots of $k$ and $\psi$ against $T(x)$ under the Gamma model indicate no obvious restrictions on the parameters based on $T(y)$, suggesting that this model is flexible enough to have generated the observed data with relatively high probability. Note that from these marginal scatterplots, it is not clear that these statistics are at all informative for the model parameters. This indicates the importance of parameterisation for visualisation, as alternatively considering method of moments estimators as summary statistics $(\hat{k},\hat{\psi})$, where $\hat{k}=\bar{x}^2/s_x^2$ and $\hat{\psi}=s_x^2/\bar{x}$, will result in strong linear relationships between $(k,\psi)$ and $(\hat{k},\hat{\psi})$. Of course, in practice direct unbiased estimators are rarely known.

\subsection{The effect of $\epsilon$}
\label{lfmcmc:section:guide:tolerance}

We now implement the LF-MCMC algorithm (Table \ref{lfmcmc:table:algorithm}) targetting the Exponential$(\lambda)$ model, with an interest in evaluating sampler performance for different $\epsilon$ values. Recall that small $\epsilon$ is required to obtain a good likelihood-free approximation to the intractable posterior $\pi_{LF}(\theta|y)\approx\pi(\theta|y)$  (see Figure \ref{lfmcmc:fig:toy}), where now $\theta=\lambda$. However, implementing  the sampler with low $\epsilon$ can be problematic in terms of initialising the chain and in achieving convergence to the stationary distribution.

An initialisation problem may occur when using weighting functions $\pi_\epsilon(y|x,\theta)$ with compact support, such as the uniform kernel (\ref{lfmcmc:eqn:uniform}) defined on $[-\epsilon,\epsilon]$. 
Here, initial chain values $(\theta_0,x_0)$ are required such that $\pi_\epsilon(y|x_0,\theta_0)\neq0$ in the denominator of the acceptance probability at time $t=1$ (Table \ref{lfmcmc:table:algorithm}). For small $\epsilon$, this is unlikely to be the case for the first such parameter vector tried. Two na\"ive strategies are to either repeatedly generate $x_0\sim\pi(x|\theta_0)$, or similarly repeatedly generate $\theta_0\sim\pi(\theta)$ and $x_0\sim\pi(x|\theta_0)$, until $\pi_\epsilon(y|x_0,\theta_0)\neq0$ is achieved. However, the former strategy may never terminate unless $\theta_0$ is located within a region of high posterior density. The latter strategy may never terminate if the prior is diffuse with respect to the posterior.
Relatedly, Markov chain convergence can be very slow for small $\epsilon$ when moving through regions of very low density, for which generating $x'\sim\pi(x|\theta')$ with $T(x')\approx T(y)$ is highly improbable.

One strategy to avoid these problems is to augment the target distribution from $\pi_{LF}(\theta,x|y)$ to $\pi_{LF}(\theta,x,\epsilon|y)$ \citep{bortot+cs07}, permitting a time-variable $\epsilon$ to improve chain mixing (see Section \ref{lfmcmc:section:samplers} for discussion on this and other strategies to improve chain mixing).
A simpler strategy is to implement a specified chain burn-in period, defined by   a monotonic decreasing sequence $\epsilon_{t+1}\leq\epsilon_t$, initialised with large $\epsilon_0$, for which $\epsilon_t=\epsilon$ remains constant at the desired level for $t\geq t^*$, beyond some (possibly random) time $t^*$ (e.g. \cite{peters+nsfy09}).
For example,  consider the linear sequence $\epsilon_t=\max\{\epsilon_0-ct,\epsilon\}$ for some $c>0$. However, the issue here is in determining the rate at which the sequence approaches the target $\epsilon$: if $c$ is too large, then $\epsilon_t=\epsilon$ before $(\theta_t,x_t)$ has reached a region of high density; if $c$ is too small, then the chain mixes well but is computationally expensive through a slow burn in.

One self-scaling option for the uniform weighting function (\ref{lfmcmc:eqn:uniform}) would be to define $\epsilon_0=\rho(T(x_0),T(y))$, and given the proposed pair $(\theta',x')$ at time $t$, propose a new $\epsilon$ value as
\begin{equation}
\label{lfmcmc:eqn:auto}
	\epsilon''=
	\max\{\epsilon, \min\{\epsilon',\epsilon_{t-1}\}\}
\end{equation}
where $\epsilon'=\rho(T(x'),T(y))>0$ is the distance between observed and simulated summary statistics. If the proposed pair $(\theta',x')$ are accepted then set $\epsilon_t=\epsilon''$, else set $\epsilon_t=\epsilon_{t-1}$. That is, the proposed $\epsilon''$ is dynamically defined as the smallest possible value that results in a non-zero weighting function $\pi_{\epsilon_t}(y|x',\theta')$ in the numerator of the acceptance probability, without going below the target $\epsilon$, and while decreasing monotonically.
If the proposed move to $(\theta',x')$ is accepted, the value $\epsilon''$ is accepted as the new state, else the previous value $\epsilon_{t-1}$ is retained.
Similar approaches could be taken with non-uniform weighting functions $\pi_\epsilon(y|x,\theta)$.

\begin{figure}
\begin{center}
\rotatebox{0}{\includegraphics[width=16cm]{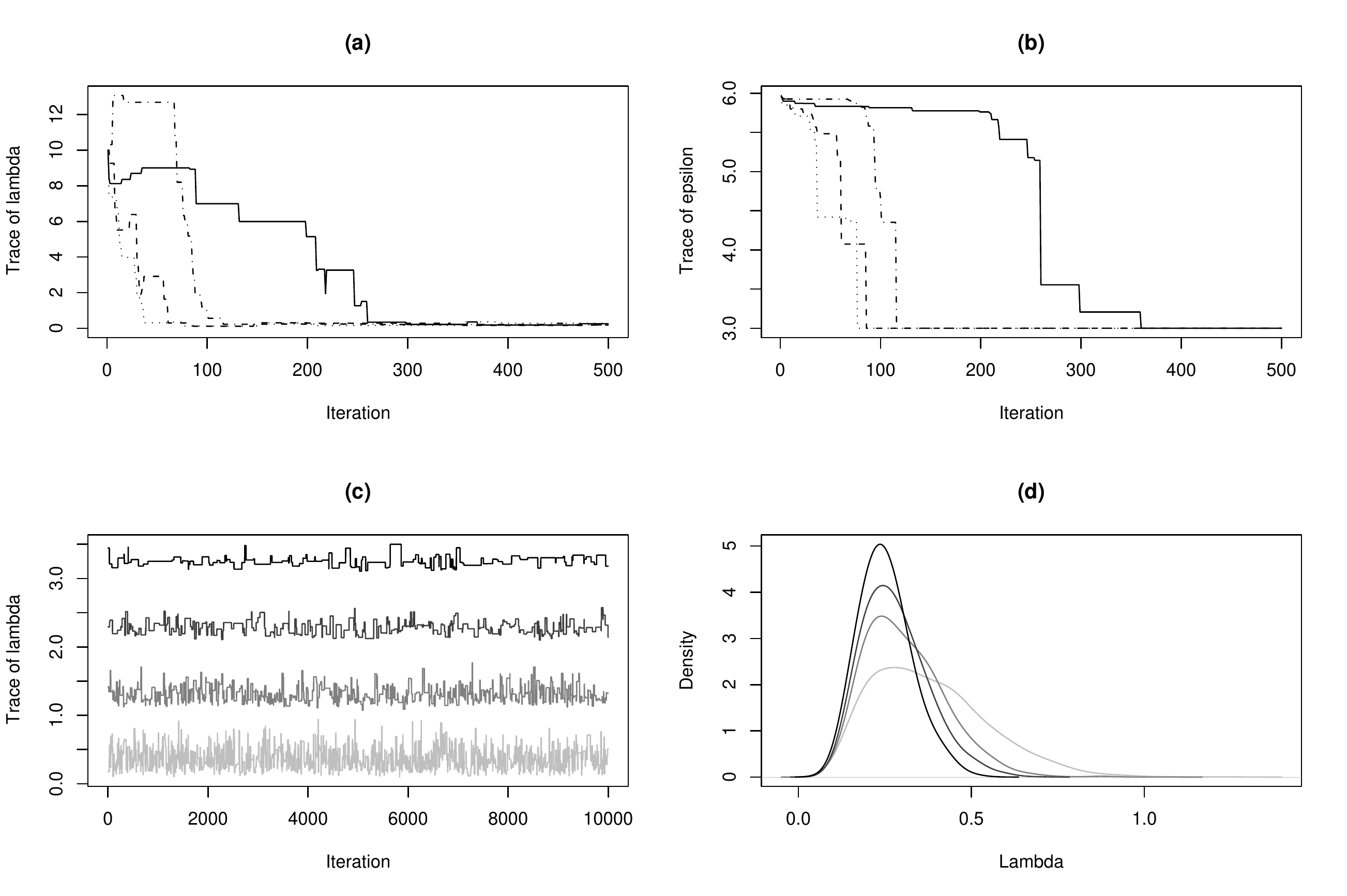}} 
 \caption{\label{lfmcmc:fig:sim:fig2}Performance of the LF-MCMC sampler for the Exponential$(\lambda)$ model. [Top plots] Trace plots of (a) $\lambda_t$ and (b) $\epsilon_t$ for four chains using the self-scaling $\{\epsilon_t\}$ sequence given by (\ref{lfmcmc:eqn:auto}). The MLE of $\lambda$ is 0.25 and the target $\epsilon$ is 3. [Bottom plots] (c) Jittered trace plots of $\lambda_t$ with different target $\epsilon=4.5$ (bottom), 4, 3.5 and 3 (top). (d) Posterior density estimates of $\lambda$ for the same chains based on a chain length of 100,000 iterations.
}
\end{center}
\end{figure}

Four trace plots of $\lambda_t$ and $\epsilon_t$ for the Exponential$(\lambda)$ model are illustrated in Figure \ref{lfmcmc:fig:sim:fig2} (a,b), using the above procedure.
All Markov chains were initialised at $\lambda_0=10$ with target $\epsilon=3$, proposals were generated via $\lambda'\sim N(\lambda_{t-1},1)$ and the distance measure 
\begin{equation}
\label{lfmcmc:eqn:mahalanobis}
	\rho(T(x),T(y))=\left\{[T(x)-T(y)]^\top\Sigma^{-1}[T(x)-T(y)]\right\}^{1/2}
\end{equation} 
is given by Mahalanobis distance. The covariance matrix $\Sigma=\mbox{Cov}(T(y))$ is estimated by the sample covariance of 1000 summary vectors $T(x)$ generated from $\pi(x|\hat{\lambda})$ conditional on $\hat{\lambda}=0.25$ the maximum likelihood estimate.
All four chains converge to the high density region at $\lambda=0.25$ quickly, although at different speeds as the sampler takes different routes through parameter space.  Mixing during burn-in  is variable between chains,
although overall convergence to $\epsilon_t=3$ is rapid. The requirement of tuning the rate of convergence, beyond specifying the final tolerance $\epsilon$, is clearly circumvented.

Figure \ref{lfmcmc:fig:sim:fig2} (c,d) also illustrates the performance of the LF-MCMC sampler, post-convergence, based on four chains of length 100,000, each with different target $\epsilon$. As expected (see discussion in Section \ref{lfmcmc:section:samplers}), smaller $\epsilon$ results in lower acceptance rates. In Figure \ref{lfmcmc:fig:sim:fig2} (c), $\epsilon=4.5$ (bottom trace), 4, 3.5 and 3 (top) result in post-convergence (of $\epsilon_t$) mean acceptance rates of  12.2\%, 6.1\%, 2.9\% and 1.1\% respectively. 
Conversely, precision (and accuracy) of the posterior marginal distribution for $\lambda$ increases with decreasing $\epsilon$ as seen in Figure \ref{lfmcmc:fig:sim:fig2} (d).

In practice, a robust procedure to identify a suitable target $\epsilon$ for the likelihood-free MCMC sampler is not yet available. 
\citet{wegmann+le09} implement the LF-MCMC algorithm with a large $\epsilon$ value to enhance chain mixing, and then perform a regression-based adjustment \citep{beaumont+zb02,blum+f09} to improve the final posterior approximation.
\citet{bortot+cs07} implement the LF-MCMC algorithm targetting the augmented posterior $\pi_{LF}(\theta,x,\epsilon|y)$ (see Section \ref{lfmcmc:sec:epsaug}), and examine the changes in $\pi_{LF}^{\mathcal E}(\theta|y)=\int_{\mathcal{E}}\int_{\mathcal Y}\pi_{LF}(\theta,x,\epsilon|y)dxd\epsilon$, with $\mathcal{E}=[0,\epsilon^*]$, for varying $\epsilon^*$. The final choice of $\epsilon^*$ is the largest value for which reducing $\epsilon^*$ further produces no obvious improvement in the posterior approximation.
This procedure may be repeated manually through repeated LF-MCMC sampler implementations at different fixed $\epsilon$ values \citep{tanaka+fls06}. Regardless, in practice $\epsilon$ is often reduced as low as possible such that computation remains within acceptable limits.

\subsection{The effect of the weighting function}
\label{lfmcmc:section:guide:weight}

The optimal form of kernel weighting function $\pi_\epsilon(y|x,\theta)$ for a given analysis is unclear at present. While the uniform weighting function (\ref{lfmcmc:eqn:uniform}) is the most common in practice -- indeed, many likelihood-free methods have this kernel written directly into the algorithm (sometimes implicitly) -- it seems credible that alternative forms may offer improved posterior approximations for given computational overheads. Some support for this is available through recently observed  links between the likelihood-free posterior approximation $\pi_{LF}(\theta|y)$ and non-parametric smoothing \citep{blum09}.

Here we evaluate the effect of the weighting function $\pi_\epsilon(y|x,\theta)$ on posterior accuracy under the Exponential$(\lambda)$ model, as measured by the one-sample Kolmogorov-Smirnov distance between the likelihood-free posterior sample and the true Gamma(21,80) posterior. To provide fair comparisons, we evaluate posterior accuracy as a function of  computational overheads,  measured by the mean post-convergence acceptance rate of the LF-MCMC sampler. The following results are based on posterior samples consisting of 1000 posterior realisations obtained by recording every 1000$^{th}$ chain state, following  a 10,000 iteration burn-in period. Figures are constructed by averaging the results of 25 sampler replications under identical conditions, for a range of $\epsilon$ values.

\begin{figure}
\begin{center}
\rotatebox{0}{\includegraphics[width=16cm]{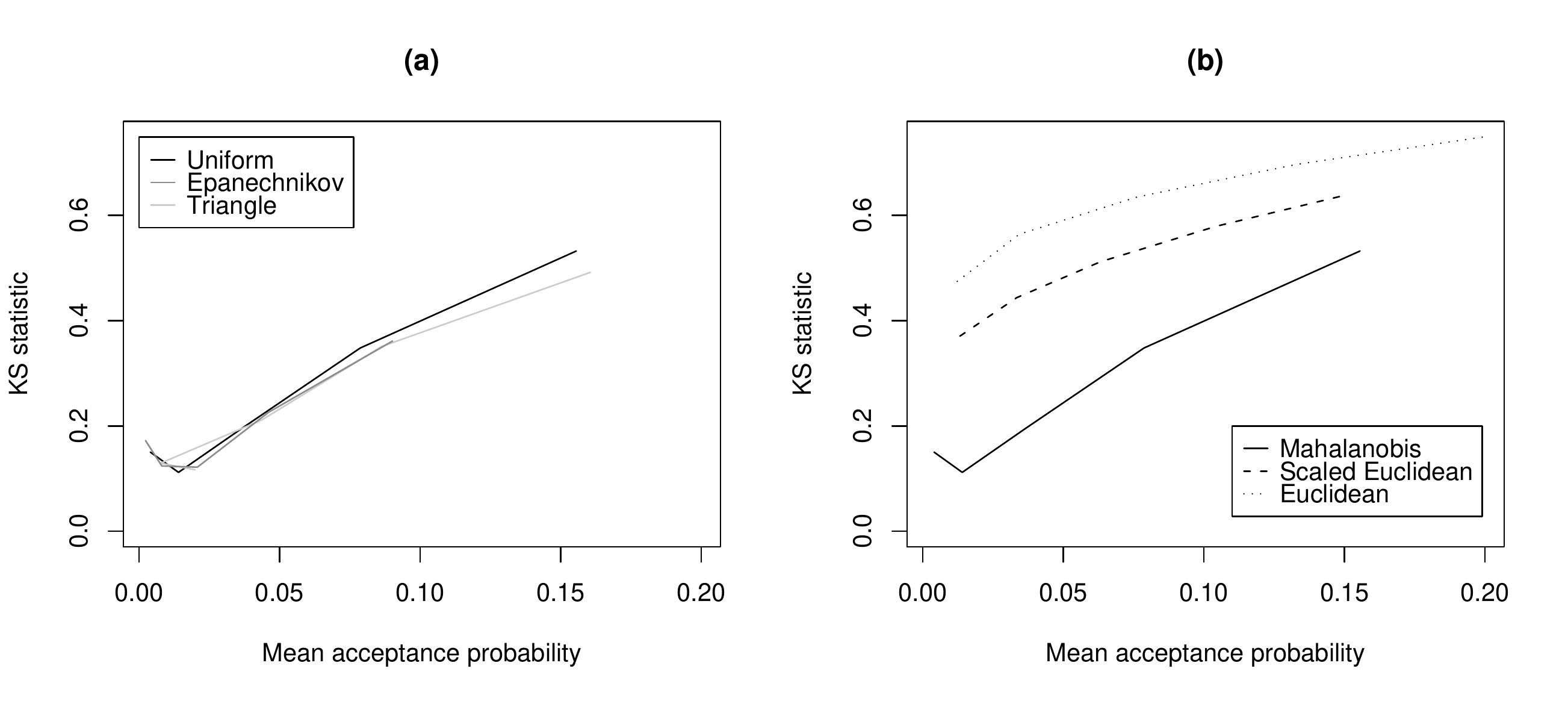}} 
 \caption{\label{lfmcmc:fig:sim:fig3}Performance of the LF-MCMC sampler for the Exponential$(\lambda)$ model under varying kernel weighting functions: (a) Mahalanobis distance between $T(x)$ and $T(y)$ evaluated on uniform, Epanechnikov and triangle kernel functions; (b) Mahalanobis, scaled Euclidean and Euclidean distance between $T(x)$ and $T(y)$ evaluated on the uniform kernel function. Sampler performance is measured  in terms of accuracy (y-axis: one-sample Kolmogorov-Smirnov test statistic evaluated between likelihood-free posterior sample and true posterior) versus computational overheads (x-axis: mean sampler acceptance probability).
}
\end{center}
\end{figure}

Figure \ref{lfmcmc:fig:sim:fig3} (a) shows the effect of varying the form of the kernel weighting function based on the Mahalanobis distance (\ref{lfmcmc:eqn:mahalanobis}). There appears little obvious difference in the accuracy of the posterior approximations in this example. However, it is credible to suspect that non-uniform weighting functions may be superior in general (e.g. \citet{blum09}; \citet{peters+fs08}). This is more clearly demonstrated in Section \ref{lfmcmc:sec:mixing}.
The slight worsening in the accuracy of the posterior approximation, indicated by the upturn  for low $\epsilon$ in Figure \ref{lfmcmc:fig:sim:fig3} (a), will be examined in more detail in Section \ref{lfmcmc:section:ss}.

Regardless of its actual form, the weighting function $\pi_\epsilon(y|x,\theta)$ should take the distribution of the summary statistics $T(\cdot)$ into consideration. \citet{fan+ps09} note that using a Euclidean distance measure (given by 
(\ref{lfmcmc:eqn:mahalanobis}) with $\Sigma=I$, the identity matrix) within (say) the uniform weighting function  (\ref{lfmcmc:eqn:uniform}),  ignores the scale and dependence (correlation) structure of $T(\cdot)$, accepting sampler moves if  $T(y)$ is within a circle of size $\epsilon$ centered on $T(x)$, rather than within an ellipse defined by $\Sigma=\mbox{Cov}(T(y))$. In theory, the form of the distance measure does not matter as in the limit $\epsilon\rightarrow0$ any effect of the distance measure $\rho$ is removed from the posterior $\pi_{LF}(\theta|y)$ i.e. $T(x)=T(y)$ regardless of the form of $\Sigma$. In practice however, with $\epsilon>0$, the distance measure can have a strong effect on the quality of the likelihood-free posterior approximation $\pi_{LF}(\theta|y)\approx\pi(\theta|y)$.

Using the uniform weighting function, Figure \ref{lfmcmc:fig:sim:fig3} (b) demonstrates the effect of using Mahalanobis distance (\ref{lfmcmc:eqn:mahalanobis}), with $\Sigma$ given by estimates of $\mbox{Cov}(T(y))$, $\mbox{diag}(\mbox{Cov}(T(y)))$ (scaled Euclidean distance) and the identity matrix $I$ (Euclidean distance). Clearly, for a fixed computational overhead (x-axis), greater accuracy is attainable by standardising and orthogonalising the summary statistics.
In this sense, Mahalanobis distance represents an approximate standardisation of the distribution of $T(y)|\tilde{\theta}$ at an appropriate point $\tilde{\theta}$ following indirect inference arguments \citep{jiang+t04}. 
As $\mathrm{Cov}(T(y))$ may vary with $\theta$, \cite{fan+ps09} suggest using an approximate MAP estimate of $\theta$, so that $\tilde{\theta}$ resides in a region of high posterior density. The assumption is then that $\mbox{Cov}(T(y))$ varies little over the region of high posterior density.

\subsection{The choice of summary statistics}
\label{lfmcmc:section:ss}

Likelihood-free computation is based on the reproduction of observed statistics $T(y)$ under the model. If the $T(y)$ are sufficient for $\theta$, then the true posterior $\pi(\theta|y)$ can be recovered exactly as $\epsilon\rightarrow 0$. If $\dim(T(y))$ is large (e.g. \cite{bortot+cs07}), then likelihood-free algorithms become computationally inefficient through the need to reproduce large numbers of summary statistics 
\citep{blum09}. 
However, low-dimensional, non-sufficient summary vectors produce less efficient estimators of $\theta$, and so generate wider posterior distributions $\pi_{LF}(\theta|y)$ than using sufficient statistics (see Section  \ref{lfmcmc:sec:example}). Ideally, low-dimensional and near-sufficient $T(y)$ are the preferred option.

Unfortunately, it is usually difficult to know which statistics are near-sufficient in practice. A brute-force strategy to address this issue is to repeat the analysis, while sequentially increasing the number of summary statistics each time (in order of their perceived importance), until no further changes to $\pi_{LF}(\theta|y)$ are observed \citep{marjoram+mpt03}. See also \citet{joyce+m08}.
If the extra statistics are {\it un}informative, the quality of approximation will remain the same, but the sampler will be less efficient.
However, simply enlarging the number of informative summary statistics is not necessarily the best way to improve the likelihood-free approximation $\pi_{LF}(\theta|y)\approx\pi(\theta|y)$, and in fact may worsen the approximation in some cases.

\begin{figure}
\begin{center}
\rotatebox{0}{\includegraphics[width=16cm]{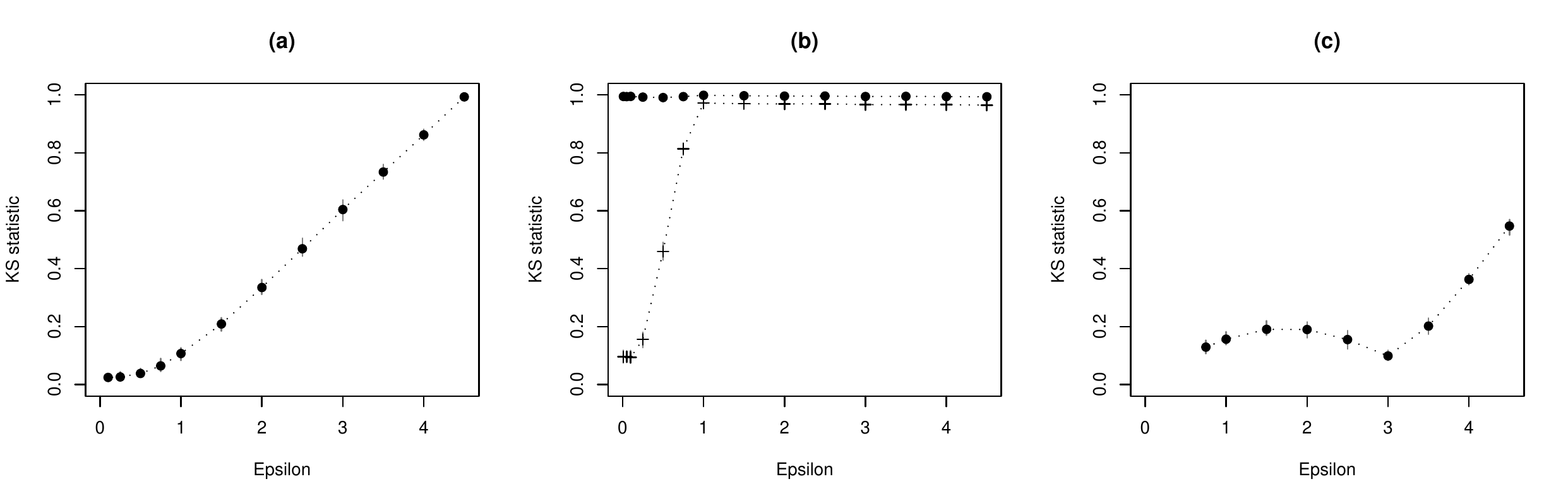}} 
 \caption{\label{lfmcmc:fig:sim:fig4}Likelihood-free posterior accuracy of the Exponential$(\lambda)$ model as a function of $\epsilon$ for differing summary statistics: (a) $T(y)=\bar{y}$; (b) $T(y)=s_y$; (c) $T(y)=(\bar{y},s_y)$. Posterior accuracy (y-axis) is measured by one-sample Kolmogorov-Smirnov (KS) test statistic evaluated between likelihood-free posterior sample and true posterior. Points and vertical lines represent KS statistic means and ranges based on 25 sampler replicates at fixed $\epsilon$ levels. Crosses in panel (b) denote KS statistic evaluated with respect to a Gamma(21,20) distribution.
}
\end{center}
\end{figure}

An example of this is provided by the present Exponential$(\lambda)$ model, where either of the two summary statistics $T(y)=(\bar{y},s_y)=(4,1)$ alone is informative for $\lambda$ (and indeed, $\bar{y}$ is sufficient), as we expect that $\lambda\approx1/\bar{y}\approx 1/s_y$ under any data generated from this model. In this respect, however, the observed values of the summary statistics provide conflicting information for the model parameter (see Section \ref{lfmcmc:sec:explore}).
Figure \ref{lfmcmc:fig:sim:fig4} examines the effect of this, by evaluating the accuracy of the likelihood-free posterior approximation $\pi_{LF}(\theta|y)\approx\pi(\theta|y)$
as a function of $\epsilon$ under different summary statistic combinations. As before, posterior accuracy is measured via the one-sample Kolmogorov-Smirnov test statistic with respect to the true Gamma(21,80)  posterior.

With $T(y)=\bar{y}$, panel (a) demonstrates that accuracy improves as $\epsilon$ decreases, as expected. 
For panel (b), with $T(y)=s_y$ (dots), the resulting $\pi_{LF}(\theta|y)$ posterior is clearly different from the true posterior for all  $\epsilon$. Of course, the limiting posterior as $\epsilon\rightarrow 0$ is (very) approximately Gamma(21,20), resulting from an Exponential model with $\lambda=1/s_y=1$, rather than Gamma(21,80) resulting from an Exponential model with $\lambda=1/\bar{y}=1/4$. The crosses in panel (b) denote the Kolmogorov-Smirnov test statistic with respect to the Gamma(21,20) distribution, which indicates that $\pi_{LF}(\theta|y)$ is roughly consistent with this distribution as $\epsilon$ decreases. That the Gamma(21,20) is not the exact limiting density (i.e. the KS statistic does not tend to zero as $\epsilon\rightarrow 0$) stems from the fact that $s_y$ is not a sufficient statistic for $\lambda$, and is less then fully efficient.

In panel (c) with $T(y)=(\bar{y},s_y)$, which contains an exactly sufficient statistic (i.e. $\bar{y}$), the accuracy of $\pi_{LF}(\theta|y)$ appears to improve with decreasing $\epsilon$, and then actually worsens before improving again. This would appear to go against the generally accepted principle,  that for sufficient statistics, decreasing $\epsilon$ will always improve the approximation $\pi_{LF}(\theta|y)\approx\pi(\theta|y)$.
Of course, the reality here is that
both of these competing statistics are pulling the likelihood free posterior in different directions, with the consequence that the limiting posterior as $\epsilon\rightarrow 0$ will be some combination of both Gamma distributions, rather than the presumed (and desired) Gamma(21,80).

This observation leads to the uncomfortable conclusion that model comparison through likelihood-free posteriors with a fixed vector of summary statistics $T(y)$, will ultimately compare distortions of those models which are overly simplified with respect to the true data generation process.
This remains true even when using sufficient statistics and for $\epsilon\rightarrow 0$.

\subsection{Improving mixing}
\label{lfmcmc:sec:mixing}

\begin{figure}
\begin{center}
\rotatebox{0}{\includegraphics[width=16cm]{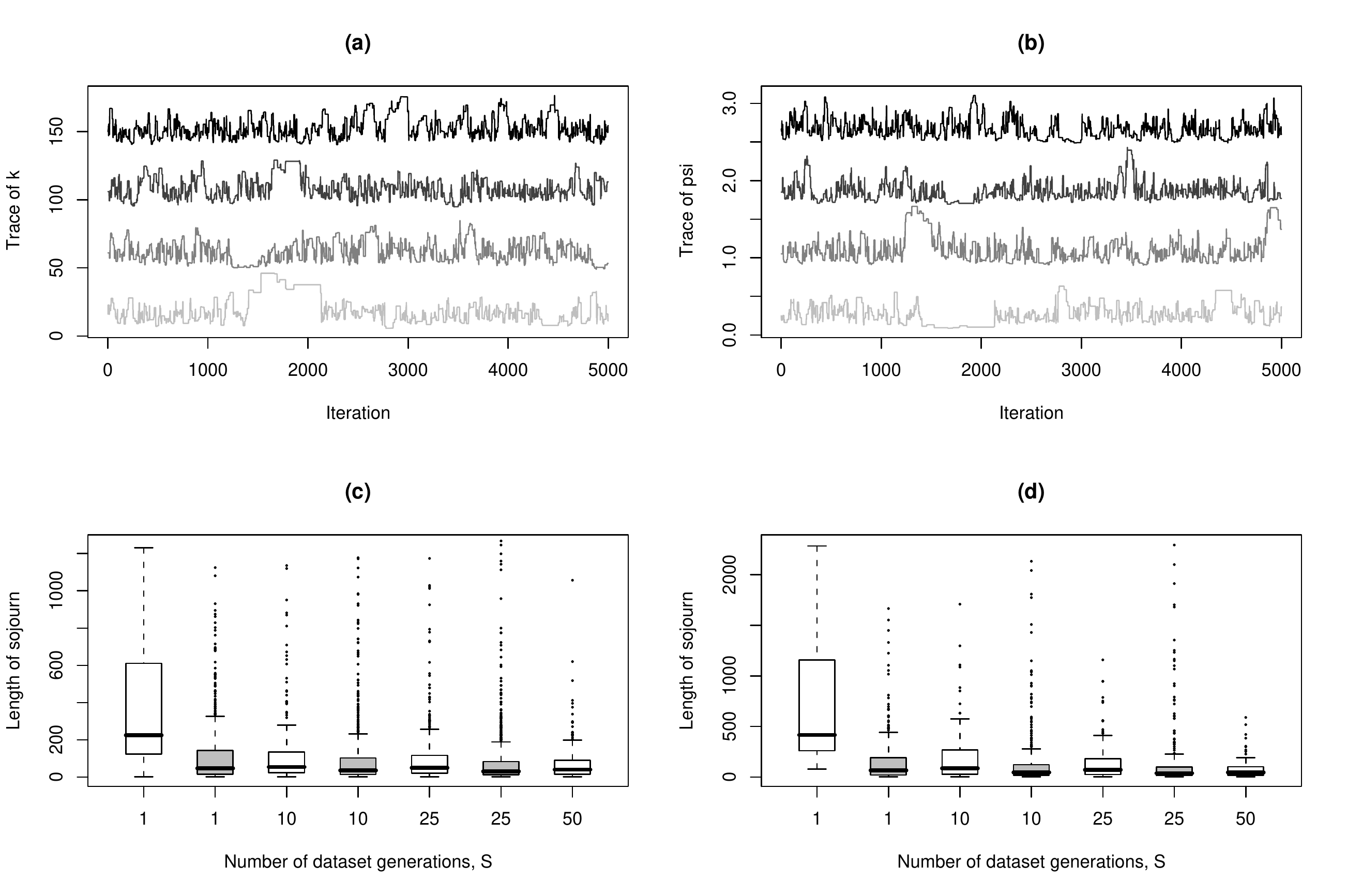}}
  \caption{\label{lfmcmc:fig:sim:sojourn}Aspects of LF-MCMC sampler performance: [Top plots] Trace plots of (a) $k$ and (b) $\psi$ parameters under the Gamma model, for varying numbers of auxiliary datasets $S=1$ (lower traces), $10, 25$ and $50$ (upper traces) using $\epsilon=2$ and the uniform kernel function $\pi_{\epsilon}(y|x,\theta)$. [Bottom plots] Distribution of sojourn lengths of parameter $k$ above (c) $\kappa=45$ and (d) $\kappa=50$ for varying numbers of auxiliary datasets. Boxplot shading indicates uniform (white) or Gaussian (grey) kernel functions $\pi_{\epsilon}(y|x,\theta)$. The Gaussian kernel sampler used $\epsilon=2/\sqrt{3}$ to ensure a comparable standard deviation with the uniform kernel sampler.
}
\end{center}
\end{figure}

Recall that 
the acceptance rate
of the LF-MCMC algorithm (Table \ref{lfmcmc:table:algorithm}) is directly related to the value of the true likelihood $\pi(y|\theta')$ at the proposed vector $\theta'$ (Section \ref{lfmcmc:section:samplers}).
While this is a necessary consequence of likelihood-free computation, it does imply poor sampler performance in regions of low probability, as the Markov chain sample-path may persist in distributional tails for long periods of time due to low acceptance probabilities \citep{sisson+ft07}.
An illustration of this shown in Figure \ref{lfmcmc:fig:sim:sojourn} (a, b: lowest light grey lines), which displays the marginal sample paths of $k$ and $\psi$ under the Gamma($k,\psi$) model, based on 5000 iterations of a sampler targetting $\pi(\theta,x|y)$ with $\epsilon=2$ and using the uniform kernel function $\pi_\epsilon(y|x,\theta)$.
At around 1400 iterations the sampler becomes stuck in the tail of the posterior for the following 700 iterations, with very little meaningful movement.

A simple strategy to improve sampler performance in this respect is to increase the number of auxiliary datasets $S$ generated under the model, either by targetting the joint posterior $\pi_{LF}(\theta,x_{1:S}|y)$ or the marginal posterior $\pi_{LF}(\theta|y)$ with $S\geq1$ Monte Carlo draws (see Section \ref{lfmcmc:sec:marginal}). This approach will reduce the variability of the acceptance probability (\ref{lfmcmc:eqn:montecarlo}), and allow the Markov chain acceptance rate to approach that of a sampler targetting the true posterior $\pi(\theta|y)$. The trace plots in Figure \ref{lfmcmc:fig:sim:sojourn} (a,b) (bottom to top) correspond to chains implementing $S=1, 10, 20$ and $50$ auxiliary dataset generations per likelihood evaluation. Visually, there is some suggestion that mixing is improved as $S$ increases. 
Note however, that for any fixed $S$, the LF-MCMC sampler may still become stuck if the sampler explores sufficiently far into the distributional tail.

Figure \ref{lfmcmc:fig:sim:sojourn} (c,d) investigates this idea from an
alternative perspective.  Based on 2 million sampler iterations, the lengths
of sojourns that the $k$ parameter spent above a fixed threshold $\kappa$ were
recorded. A sojourn length is defined as the consecutive number of iterations in which
the parameter $k$ remains above $\kappa$.
Intuitively, if likelihood-free samplers tend to persist in
distributional tails, the length of the sojourns will be much larger for the worse performing samplers. 
Figure \ref{lfmcmc:fig:sim:sojourn} (c,d) shows the distributions of sojourn lengths for samplers with $S=1, 10, 25$ and $50$ auxiliary datasets, with $\kappa=45$ (panel c) and $\kappa=50$ (panel d). Boxplot shading indicates use of the uniform (white) or Gaussian (grey) weighting function $\pi_{\epsilon}(y|x,\theta)$.

A number of points are immediately apparent. Firstly, chain mixing is poorer the further into the tails the sampler explores. This is illustrated by the increased scale of the sojourn lengths for $\kappa=50$ compared to $\kappa=45$.
Secondly, increasing $S$ by a small amount substantially reduces chain tail persistence. As $S$ increases further, the Markov chain performance approaches that of a sampler directly targetting the true posterior $\pi(\theta|y)$, and so less performance gains are observed by increasing $S$ beyond a certain point.
Finally, there is strong evidence to suggest that LF-MCMC algorithms using weighting kernel functions $\pi_\epsilon(y|x,\theta)$ that do not generate large numbers of zero-valued likelihoods will possess superior performance to those which do. Here use of the Gaussian weighting kernel clearly outperforms the uniform kernel in all cases.
In summary, it would appear that the choice of kernel weighting function $\pi_{\epsilon}(\theta|y)$ has a larger impact on sampler performance than the number of auxiliary datasets $S$.

\subsection{Evaluating model mis-specification}

\begin{figure}
\begin{center}
\rotatebox{0}{\includegraphics[width=16cm]{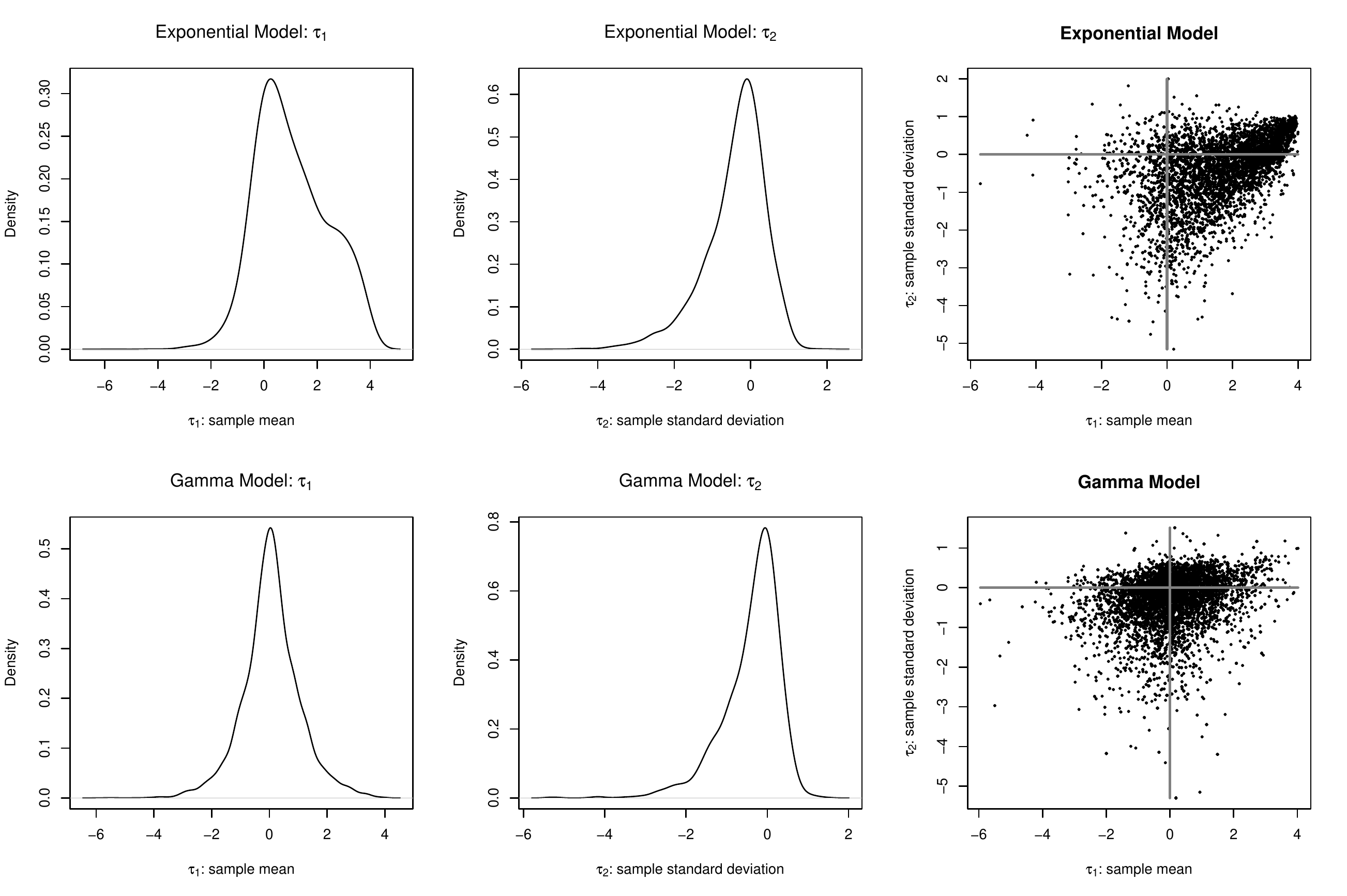}} 
 \caption{\label{lfmcmc:fig:ratmann} Marginal likelihood-free posterior distributions $\pi_{LF}(\tau|y)$ of the error-distribution augmented model (\ref{lfmcmc:eqn:ratmann}), under the Exponential (top plots) and Gamma (bottom plots) models. Plots are based on 50,000 sampler iterations.
}
\end{center}
\end{figure}

In order to evaluate the adequacy of both Exponential and Gamma models in terms of their support for the observed data $T(y)=(\bar{y},s_y)$, we fit the error-distribution augmented model (\ref{lfmcmc:eqn:ratmann}) given by
\[
	\pi_{LF}(\theta, x_{1:S}, \tau|y)
	 := 
	\min_r \hat{\xi}_r(\tau_r|y, x_{1:S}, \theta) \pi(x_{1:S}|\theta) \pi(\theta)\pi(\tau),
\]
as described in Section \ref{lfmcmc:sec:epsaug} \citep{ratmann+ahwr09}. The vector $\tau=(\tau_1,\tau_2)$ with $\tau_r=T_r(x)-T_r(y)$ for $r=1,2$, describes the error under the model in reproducing the observed summary statistics $T(y)$.
The marginal likelihood-free posterior $\pi_{LF}(\tau|y)$ should be centered on the zero vector for models which can adequately account for the observed data.

We follow \citet{ratmann+ahwr09} in specifying $K$  in (\ref{lfmcmc:eqn:ratmannK}) as a biweight (quartic) kernel with an adaptive bandwidth $\epsilon_r$ determined by twice the interquartile range of $T_r(x^s)-T_r(y)$ given $x_{1:S}=(x^1,\ldots,x^S)$. The prior on the error $\tau$ is determined as $\pi(\tau)=\prod_r\pi(\tau_r)$, where $\pi(\tau_r)=\exp(-|\tau_r|/\delta_r)/(2\delta_r)$ with $\delta_1=\delta_2=0.75$ for both Exponential and Gamma models.

Based on 50,000 sampler iterations using $S=50$ auxiliary datasets, the resulting bivariate posterior $\pi_{LF}(\tau|y)$ is illustrated in Figure \ref{lfmcmc:fig:ratmann} for both models.
From these plots, the errors $\tau$ under the Gamma model (bottom plots) are clearly centered on the origin, with 50\% marginal high-density regions given by $\tau_1|y\sim[-0.51, 0.53]$ and $\tau_2|y\sim[-0.44,0.22]$ \citep{ratmann+ahwr09}. 
However for the Exponential model (top plots), while the marginal 50\% high density regions $\tau_1|y\sim[-0.32, 1.35]$ and $\tau_2|y\sim[-0.55,0.27]$ also both contain zero, there is some indication of model mis-specification as the joint posterior error distribution $\tau|y$ is not fully centered on the zero vector. Based on this assessment, and recalling the discussion on the exploratory analysis in Section 
\ref{lfmcmc:sec:explore}, the Gamma model would appear to provide a better overall fit to the observed data.

\section{Discussion}

In the early 1990's, the introduction of accessible Markov chain Monte Carlo samplers provided the catalyst for a rapid adoption of Bayesian methods and inference as credible tools in model-based research. Twenty years later, the demand for computational techniques capable of handling the types of models inspired by complex hypotheses has resulted in new classes of simulation-based inference, that are again expanding the applicability and relevance of the Bayesian paradigm to new levels.

While the focus of the present article centers on Markov chain-based, likelihood-free simulation, alternative  methods to obtain samples from $\pi_{LF}(\theta|y)$ have been developed, each with their own benefits and drawbacks.
While MCMC-based samplers can be more efficient than rejection sampling algorithms, the tendency of sampler performance to degrade in regions of low posterior density (see Section \ref{lfmcmc:sec:mixing}; \citet{sisson+ft07}) can be detrimental to sampler efficiency.
One class of methods, based on the output of a rejection sampler with a high $\epsilon$ value (for efficiency), uses standard multivariate regression methods to estimate the relationship between the summary statistics $T(x)$ and parameter vectors $\theta$ \citep{beaumont+zb02,blum+f09,marjoram+t06}. The idea is then to approximately transform the sampled observations from $(\theta,T(x))$ to $(\theta^*,T(y))$ so that the adjusted likelihood-free posterior $\pi_{LF}(\theta,x|y)\rightarrow\pi_{LF}(\theta^*,y|y)\approx\pi(\theta|y)$ is an improved approximation.
Further attempts to improve sampler efficiency over MCMC-based methods have resulted in the development of likelihood-free sequential Monte Carlo and sequential importance sampling algorithms \citep{sisson+ft07,peters+fs08,beaumont+cmr09, toni+wsis09,delmoral+dj08}. Several authors have reported that likelihood-free sequential Monte Carlo approaches can outperform their MCMC counterparts \citep{mckinley+cd09,sisson+ft07}.

There remain many open research questions in likelihood-free Bayesian inference. These include how to select and incorporate the vectors of summary statistics $T(\cdot)$, how to perform posterior simulation in the most efficient manner, and which form of joint likelihood-free posterior models and kernel weighting functions admit the most effective marginal approximation to the true posterior $\pi_{LF}(\theta|y)\approx\pi(\theta|y)$. Additionally, the links to existing bodies of research, including non-parametrics \citep{blum09} and indirect inference \citep{jiang+t04}, are at best poorly understood.

Finally, there is an increasing trend towards using likelihood-free inference for model selection purposes \citep{grelaud+rmrt09,toni+wsis09}. While this is a natural extension of inference for individual models, the analysis in Section \ref{lfmcmc:section:ss} urges caution and suggests that further research is needed into the effect of the likelihood-free approximation both within models and on the marginal likelihoods $\pi_{LF}(y)=\int_{{\mathcal Y}}\pi_{LF}(\theta|y)d\theta$ upon which model comparison is based.

%
\section*{Acknowledgments}
This work was supported by the Australian Research Council through the Discovery Project scheme (DP0664970 and DP1092805).

%
\bibliographystyle{apalike} 
\bibliography{lfmcmc}

\begin{thebibliography}{}

\bibitem[Andrieu et~al., 2008]{andrieu+bdr08}
Andrieu, C., Berthelsen, K.~K., Doucet, A., and Roberts, G.~O. (2008).
\newblock The expected auxiliary variable method for {Monte Carlo} simulation.
\newblock Technical report, In preparation.

\bibitem[Beaumont et~al., 2009]{beaumont+cmr09}
Beaumont, M.~A., Cornuet, J.-M., Marin, J.-M., and Robert, C.~P. (2009).
\newblock Adaptive approximate {Bayesian} computation.
\newblock {\em Biometrika}, in press.

\bibitem[Beaumont et~al., 2002]{beaumont+zb02}
Beaumont, M.~A., Zhang, W., and Balding, D.~J. (2002).
\newblock Approximate {Bayesian} computation in population genetics.
\newblock {\em Genetics}, 162:2025 -- 2035.

\bibitem[Blum, 2009]{blum09}
Blum, M. G.~B. (2009).
\newblock Approximate {Bayesian} computation: a non-parametric perspective.
\newblock Technical report, Universit\'e Joseph Fourier, Grenoble, France.

\bibitem[Blum and Francois, 2009]{blum+f09}
Blum, M. G.~B. and Francois, O. (2009).
\newblock Non-linear regression models for approximate {Bayesian} computation.
\newblock {\em Statistics and Computing}, page in press.

\bibitem[Blum and Tran, 2009]{blum+t09}
Blum, M. G.~B. and Tran, V.~C. (2009).
\newblock {HIV} with contact-tracing: {A} case study in approximate {Bayesian}
  computation.
\newblock Technical report, Universit\'e Joseph Fourier.

\bibitem[Bortot et~al., 2007]{bortot+cs07}
Bortot, P., Coles, S.~G., and Sisson, S.~A. (2007).
\newblock Inference for stereological extremes.
\newblock {\em Journal of the American Statistical Association}, 102:84--92.

\bibitem[{Del Moral} et~al., 2008]{delmoral+dj08}
{Del Moral}, P., Doucet, A., and Jasra, A. (2008).
\newblock Adaptive sequential {Monte Carlo} samplers.
\newblock Technical report, University of Bordeaux.

\bibitem[Drovandi and Pettitt, 2009]{drovandi+p09}
Drovandi, C.~C. and Pettitt, A.~N. (2009).
\newblock A note on {Bayesian} estimation of quantile distributions.
\newblock Technical report, Queensland University of Technology.

\bibitem[Fagundes et~al., 2007]{fagundes+rbnsbe07}
Fagundes, N. J.~R., Ray, N., Beaumont, M.~A., Neuenschwander, S., Salzano,
  F.~M., Bonatto, S.~L., and Excoffier, L. (2007).
\newblock Statistical evaluation of alternative models of human evolution.
\newblock {\em Proc. Natl. Acad. Sci. USA}, 104:17614--17619.

\bibitem[Fan et~al., 2010]{fan+ps09}
Fan, Y., Peters, G.~W., and Sisson, S.~A. (2010).
\newblock Impoved efficiency in approximate {Bayesian} computation.
\newblock Technical report, University of New South Wales.

\bibitem[Geyer and Thompson, 1995]{geyer+t95}
Geyer, C.~J. and Thompson, E.~A. (1995).
\newblock Annealing {Markov chain Monte Carlo} with applications to ancestral
  inference.
\newblock {\em Journal of the American Statistical Association}, 90:909--920.

\bibitem[Grelaud et~al., 2009]{grelaud+rmrt09}
Grelaud, A., Robert, C.~P., Marin, J.-M., Rodolphe, F., and Taly, J.-F. (2009).
\newblock {ABC} likelihood-free methods for model choice in gibbs random
  fields.
\newblock {\em Bayesian Analysis}, 4:317--336.

\bibitem[Hamilton et~al., 2005]{hamilton+crhbe05}
Hamilton, G., Currat, M., Ray, N., Heckel, G., Beaumont, M.~A., and Excoffier,
  L. (2005).
\newblock Bayesian estimation of recent migration rates after a spatial
  expansion.
\newblock {\em Genetics}, 170:409--417.

\bibitem[Jabot and Chave, 2009]{jabot+c09}
Jabot, F. and Chave, J. (2009).
\newblock Inferring the parameters of the netural theory of biodiversity using
  phylogenetic information and implications for tropical forsts.
\newblock {\em Ecology Letters}, 12:239--248.

\bibitem[Jiang and Turnbull, 2004]{jiang+t04}
Jiang, W. and Turnbull, B. (2004).
\newblock The indirect method: {Inference} based on intermediate statistics --
  {A} synthesis and examples.
\newblock {\em Statistical Science}, 19:238--263.

\bibitem[Joyce and Marjoram, 2008]{joyce+m08}
Joyce, P. and Marjoram, P. (2008).
\newblock Approximately sufficient statistics and {Bayesian} computation.
\newblock {\em Statistical Applications in Genetics and Molecular Biology},
  7(1):no. 23.

\bibitem[Luciani et~al., 2009]{luciani+sjft09}
Luciani, F., Sisson, S.~A., Jiang, H., Francis, A., and Tanaka, M.~M. (2009).
\newblock The high fitness cost of drug resistance in mycobacterium
  tuberculosis.
\newblock {\em Proc. Natl. Acad. Sci. USA}, 106:14711--14715.

\bibitem[Marjoram et~al., 2003]{marjoram+mpt03}
Marjoram, P., Molitor, J., Plagnol, V., and Tavar\'e, S. (2003).
\newblock Markov chain {Monte Carlo} without likelihoods.
\newblock {\em Proc. Natl. Acad. Sci. USA}, 100:15324 -- 15328.

\bibitem[Marjoram and Tavar\'e, 2006]{marjoram+t06}
Marjoram, P. and Tavar\'e, S. (2006).
\newblock Modern computational approaches for analysing molectular genetic
  variation data.
\newblock {\em Nature Reviews: Genetics}, 7:759--770.

\bibitem[McKinley et~al., 2009]{mckinley+cd09}
McKinley, T., Cook, A.~R., and Deardon, R. (2009).
\newblock Inference in epidemic models without likelihoods.
\newblock {\em The International Journal of Biostatistics}, 5: article 24.

\bibitem[Neal, 2003]{neal03}
Neal, R.~M. (2003).
\newblock Slice sampling.
\newblock {\em Annals of Statistics}, 31:705--767.

\bibitem[Nevat et~al., 2008]{nevat+py08}
Nevat, I., Peters, G.~W., and Yuan, J. (2008).
\newblock Coherent detection for cooperative networks with arbitrary relay
  functions using likelihood-free inference.
\newblock Technical report, University of New South Wales.

\bibitem[Peters et~al., 2008]{peters+fs08}
Peters, G.~W., Fan, Y., and Sisson, S.~A. (2008).
\newblock On sequential {Monte Carlo}, partial rejection control and
  approximate {Bayesian} computation.
\newblock Technical report, University of New South Wales.

\bibitem[Peters and Sisson, 2006]{peters+s06}
Peters, G.~W. and Sisson, S.~A. (2006).
\newblock Bayesian inference, {Monte Carlo} sampling and operational risk.
\newblock {\em Journal of Operational Risk}, 1(3).

\bibitem[Peters et~al., 2009]{peters+sf09}
Peters, G.~W., Sisson, S.~A., and Fan, Y. (2009).
\newblock Likelihood-free {Bayesian} models for $\alpha$-stable models.
\newblock Technical report, University of New South Wales.

\bibitem[Pettitt, 2006]{pettitt06}
Pettitt, A.~N. (2006).
\newblock Isaac Newton Institute Workshop on MCMC, Cambridge UK, 30 October - 2
  November, 2006.

\bibitem[Pritchard et~al., 1999]{pritchard+spf99}
Pritchard, J.~K., Seielstad, M.~T., Perez-Lezaun, A., and Feldman, M.~W.
  (1999).
\newblock Population growth of human {Y} chromosomes: {A} study of {Y}
  chromosome microsatellites.
\newblock {\em Molecular Biology and Evolution}, 16:1791--1798.

\bibitem[Ratmann et~al., 2009]{ratmann+ahwr09}
Ratmann, O., Andrieu, C., Hinkley, T., Wiuf, C., and Richardson, S. (2009).
\newblock Model criticism based on likelihood-free inference, with an
  application to protein network evolution.
\newblock {\em Proc. Natl. Acad. Sci. USA}, 106:10576--10581.

\bibitem[Ratmann et~al., 2007]{ratmann+jhsrw07}
Ratmann, O., Jorgensen, O., Hinkley, T., Stumpf, M., Richardson, S., and Wiuf,
  C. (2007).
\newblock Using likelihood-free inference to compare evolutionary dynamics of
  the protien networks of h. pylori and p. falciparum.
\newblock {\em PLoS Comp. Biol.}, 3:e230.

\bibitem[Reeves and Pettitt, 2005]{reeves+p05}
Reeves, R.~W. and Pettitt, A.~N. (2005).
\newblock A theoretical framework for approximate {Bayesian} computation.
\newblock In Francis, A.~R., Matawie, K.~M., Oshlack, A., and Smyth, G.~K.,
  editors, {\em Proceedings of the 20th International Workshop for Statistical
  Modelling, Sydney Australia, July 10-15, 2005}, pages 393--396.

\bibitem[Sisson et~al., 2007]{sisson+ft07}
Sisson, S.~A., Fan, Y., and Tanaka, M.~M. (2007).
\newblock {Sequential Monte Carlo} without likelihoods.
\newblock {\em Proc. Natl. Acad. Sci.}, 104:1760--1765. Errata (2009),
  106:16889.

\bibitem[Sisson et~al., 2008]{sisson+pfb08}
Sisson, S.~A., Peters, G.~W., Fan, Y., and Briers, M. (2008).
\newblock Likelihood-free samplers.
\newblock Technical report, University of New South Wales.

\bibitem[Tanaka et~al., 2006]{tanaka+fls06}
Tanaka, M.~M., Francis, A.~R., Luciani, F., and Sisson, S.~A. (2006).
\newblock Using {Approximate Bayesian Computation} to estimate tuberculosis
  transmission parameters from genotype data.
\newblock {\em Genetics}, 173:1511--1520.

\bibitem[Tavar\'e et~al., 1997]{tavare+bgd97}
Tavar\'e, S., Balding, D.~J., Griffiths, R.~C., and Donnelly, P. (1997).
\newblock Inferring coalescence times from {DNA} sequence data.
\newblock {\em Genetics}, 145(505-518).

\bibitem[Tierney and Mira, 1999]{tierney+m99}
Tierney, L. and Mira, A. (1999).
\newblock Some adaptive {Monte Carlo} methods for {Bayesian} inference.
\newblock {\em Statistics in medicine}, 18:2507 -- 15.

\bibitem[Toni et~al., 2009]{toni+wsis09}
Toni, T., Welch, D., Strelkowa, N., Ipsen, A., and Stumpf, M. P.~H. (2009).
\newblock Approximate {Bayesian} computation scheme for parameter inference and
  model selection in dynamical systems.
\newblock {\em J. R. Soc. Interface}, 6:187--202.

\bibitem[Wegmann et~al., 2009]{wegmann+le09}
Wegmann, D., Leuenberger, C., and Excoffier, L. (2009).
\newblock Efficient approximate {Bayesian} computation coupled with {Markov}
  chain {Monte Carlo} without likelihood.
\newblock {\em Genetics}, 182:1207--1218.

\bibitem[Wilkinson, 2008]{wilkinson08}
Wilkinson, R.~D. (2008).
\newblock Approximate {Bayesian} computation ({ABC}) gives exact results under
  the assumption of model error.
\newblock Technical report, Dept. of Probability and Statistics, University of
  Sheffield.

\bibitem[Wilkinson and Tavar\'e, 2009]{wilkinson+t09}
Wilkinson, R.~D. and Tavar\'e, S. (2009).
\newblock Estimating primate divergence times by using conditioned
  birth-and-death processes.
\newblock {\em Theoretical Population Biology}, 75:278--285.

\end{thebibliography}

\end{doublespace} 
\end{document}